\documentclass{emulateapj}
\usepackage{longtable, multirow}
\usepackage{rotating}
\bibpunct{[}{]}{,}{n}{}{;}

\newcommand{\lax }{{\lower0.8ex\hbox{$\buildrel <\over\sim$}}}
\newcommand{\gax }{{\lower0.8ex\hbox{$\buildrel >\over\sim$}}}
\newcommand{\nhintr}{\ifmmode{ N_{H}^{intr}} \else N$_{H}^{intr}$\fi}
\newcommand{\nhgal}{\ifmmode{ N_{H}^{Gal}} \else N$_{H}^{Gal}$\fi}
\newcommand{\Lbol}{\ifmmode{L_{\rm Bol}} \else $L_{\rm Bol}$\fi}
\newcommand{\logLx}{\ifmmode{{\rm log}\,L_X} \else log\,$L_X$\fi}
\newcommand{\LEdd}{\ifmmode{L_{\rm Edd}} \else $L_{\rm Edd}$\fi}
\newcommand{\REdd}{\ifmmode{L_{\rm Bol}/L_{\rm Edd}} \else $L_{\rm Bol}/L_{\rm Edd}$\fi}
\newcommand{\aox}{\ifmmode{\alpha_{\rm ox}} \else $\alpha_{\rm ox}$\fi} 

\begin{document}

\shorttitle{Empirical Links between XRB and AGN Accretion} 
\shortauthors{Trichas et al.}
\received{2013}
\title{Empirical Links between XRB and AGN accretion using\\ the
  complete z$<$0.4 spectroscopic CSC/SDSS Catalog} 
\author{Markos Trichas\altaffilmark{1,2}}
\email{mtrichas@cfa.harvard.edu}
\author{Paul J. Green\altaffilmark{1}, 
Anca Constantin\altaffilmark{3},
Tom Aldcroft\altaffilmark{1},
Malgosia Sobolewska\altaffilmark{1},
Ashley K. Hyde\altaffilmark{4},
Hongyan Zhou\altaffilmark{5},
Dong-Woo Kim\altaffilmark{1},
Daryl Haggard\altaffilmark{6},
Brandon C. Kelly\altaffilmark{7},
Eleni Kalfountzou \altaffilmark{8}}
\altaffiltext{1}{Harvard-Smithsonian Center for Astrophysics,
  Cambridge, MA 02138, USA}
 \altaffiltext{2}{EADS Astrium, Gunnels Wood Road, Stevenage, SG1 2AS, UK}
\altaffiltext{3}{Department of Physics and Astronomy, James Madison
  University, PHCH, Harrisonburg, VA 22807, USA}
\altaffiltext{4}{Astrophysics Group, Imperial College London, London SW7 2AZ, UK}
\altaffiltext{5}{Center for Astrophysics, University of Science and Technology of China, Hefei 230026, China}
\altaffiltext{6}{Center for Interdisciplinary Exploration and Research
  in Astrophysics, Northwestern University, 2145 Sheridan Road,
  Evanston, IL 60208, USA}
\altaffiltext{7}{Department of Physics, Broida Hall, University of California, Santa Barbara, CA, 93107, USA}
\altaffiltext{8}{Center for Astrophysics, Science $\&$ Technology Research Institute, University of Hertfordshire, Hatfield, AL10 9AB, UK}

\begin{abstract}
Striking similarities have been seen between accretion signatures of
Galactic X-ray binary (XRB) systems and active galactic nuclei (AGN).
XRB spectral states show a V-shaped correlation between X-ray spectral
hardness and Eddington ratio as they vary, and some AGN samples reveal
a similar trend, implying analogous processes at vastly larger masses
and timescales.  To further investigate the analogies, we have matched
617 sources from the Chandra Source Catalog to SDSS spectroscopy, and
uniformly measured both X-ray and optical spectral characteristics
across a broad range of AGN and galaxy types. We provide useful
tabulations of X-ray spectral slope for broad and narrow line AGN,
star-forming and passive galaxies and composite systems, also updating
relationships between optical (H$\alpha$ and [O\,III]) line emission and
X-ray luminosity.  We further fit broadband spectral energy
distributions with a variety of templates to estimate bolometric
luminosity.  Our results confirm a significant trend in AGN between
X-ray spectral hardness and Eddington ratio expressed in X-ray
luminosity, albeit with significant dispersion.  The trend is not
significant when expressed in the full bolometric or
template-estimated AGN luminosity.  We also confirm a
relationship between the X-ray/optical spectral slope \aox\, 
and Eddington ratio, but it may not follow the trend predicted by 
analogy with XRB accretion states. 
\end{abstract}


\section{Introduction}
There  is now  strong evidence  that powerful  active  galactic nuclei
(AGN) play a key role  in the evolution of galaxies. The correlation
of central  black hole (M$_{\rm BH}$)  and stellar bulge  mass
(e.g. Gultekin et al. 2009; 2012), and  the similarity between  the
cosmic star  formation history 
(e.g. Hopkins  $\&$ Beacom 2006)  and cosmic  M$_{\rm BH}$ assembly
history  (e.g.  Aird  et  al 2010)  both  suggest that  the growth  of
supermassive  black holes  (SMBH) is  related  to the  growth of  host
galaxies.  Understanding what  drives the  formation and  co-evolution of
galaxies and their  central SMBHs remains one of  the most significant
challenges   in  extragalactic   astrophysics. Understanding the
feedback mechanisms, hence the AGN energy production, remains a
fundamental question that needs to be answered. 

Recent attention has focused on models where AGN feedback   
regulates the star formation in the host galaxy. These scenarios 
are consistent with the $M_{\rm BH}$ - $\sigma$ relation and make various
predictions for AGN properties, including the environmental dependence
of the AGN/galaxy interplay and the relative timing of periods of peak
star formation and nuclear accretion activity. The key feature of
these  models is that they can potentially link the apparently
independent observed relations between star formation, AGN activity
and large  scale structure to the same underlying physical
process. For example, in the ``radio-mode" model of Croton et
al. (2006), accretion of gas from cooling flows in dense environments
(e.g. group,  cluster) may produce relatively low-luminosity AGN,
which in  turn heat the bulk of the cooling gas and prevent it from
falling  into the galaxy center to form stars. Alternatively, Hopkins
et al.  (2006) propose that mergers trigger luminous QSOs and
circumnuclear starbursts, which both feed and obscure the central
engine for most of its active lifetime. In this scenario, AGN outflows
eventually sweep away the dust and gas clouds, thereby quenching the
star formation. This ``QSO-mode" likely dominates in poor environments
(e.g. field, group), as the high-velocity encounters, common in dense
regions, do not favour mergers. These proposed models make clear,
testable predictions about the  properties of AGN, while observational
constraints provide first-order confirmation of this theoretical
picture (e.g. Trichas  et al. 2009; 2012). Merger-driven scenarios for
example, predict an association between optical morphological
disturbances,  star formation and an intense obscured AGN phase in low
density regions. The ``radio mode" model,  in contrast, invokes  
milder AGN activity in early-type hosts and relatively dense
environments with little or no star formation.

Low-redshift galaxies offer the best observational testbeds to study quasar evolution.  While environmental studies of nearby AGN are
consistent with non- merger-driven fueling (Constantin $\&$ Vogeley
2006; Constantin, Hoyle, $\&$ Vogeley 2008), analysis of the observed
distribution of Eddington ratios as a function of the BH masses
suggest that at z$\sim$0 there might be two distinct regimes of BH
growth, which are determined by the supply of cold gas in the host
bulge (Kauffmann $\&$ Heckman 2009).  Optical studies of narrow
emission line galaxies using emission line ratio diagnostics
(e.g. Trichas et al. 2010; Kalfountzou et al. 2011) although quite
successful in identifying cases where the dominant mechanism is either
accretion onto a black hole or radiation from hot young stars, remain
inconclusive for LINERs and composite objects. However, using the latter method,  Constantin et
al. (2009) revealed a sequence from SF via AGN
to quiescence which may be the first empirical evidence for a duty cycle 
analogous to that of the high-z quasars.

The X-ray emission arguably affords, the most sensitive test for
measurements of the intensity and efficiency of accretion. Combining
X-ray properties with optical emission line ratios for a large
unbiased sample of low redshift galaxies can be especially useful
because of the high quality diagnostics available.  
The Chandra Source Catalog (Evans et al. 2010) when cross-matched with
the SDSS, provides 
an unprecedented number of galaxies in the local Universe for which we
can combine measurements of both the X-ray and optical
emission. Previous studies of the relation between the X-ray nuclear
emission, optical emission line activity and black hole masses provide
important physical constraints to the AGN accretion. The conclusions
are that low-luminosity AGN (LLAGN) are probably scaled down versions
of more luminous AGN (e.g., Panessa et al. 2006), and that $M_{\rm BH}$ is
not the main driver of the X-ray properties (Greene $\&$ Ho 2007). The
LLAGN are claimed to be X-ray detected at relatively high rates, and
are found to be relatively unabsorbed (e.g. Miniutti et al. 2009),
with the exception of those known to be Compton thick. Nonetheless,
the X-ray investigations of AGN activity at its lowest levels remain
largely restricted to LINERs and Seyferts.

\begin{figure} 
\begin{center}
   \begin{minipage}[c]{9.2 cm}
        \includegraphics[width=1.0\textwidth]{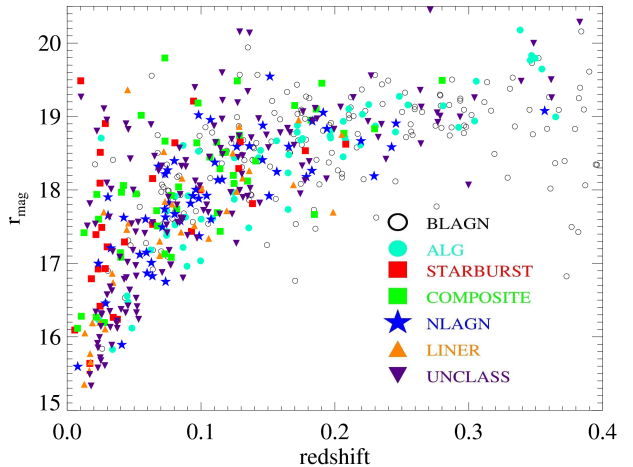}
   \end{minipage}
\end{center}
\caption{Dereddened SDSS $r$ band mag (modelMag) plotted against
redshift for our full $z<0.4$ spectroscopic sample detected in the
Chandra Source Catalog v1.1.  Black open circles are broad line AGN,
cyan filled circles are absorption line galaxies (ALG), blue stars are
narrow-line AGN (NLAGN), red squares are star-forming galaxies
(STARBURST), orange triangles are low ionization narrow emission line
region galaxies (LINERs), green squares are composite systems, and
magenta downward triangles are unclassified sources.}
\label{zr}
\end{figure}

In this work, we utilize the largest ever sample of 
galaxies with available optical spectroscopy and X-ray detections,
a total of 617 sources, to build on the work done by Constantin et
al. (2009). We combine the CSC X-ray detections with a sample of SDSS
DR7 spectroscopically identified nearby galaxies that includes broad
line objects, creating a large sample of galaxy nuclei that spans a
range of optical spectral types, from absorption line (passive) to
actively line emitting systems, including the star-forming and
actively accreting types, along with those of mixed or ambiguous
ionization. Our main goal is try to verify whether we see the same turning point found by both Constantin et al. (2009) and Wu $\&$ Gu (2008)in the $\Gamma$ - L\L$_{Edd}$ relation that occurs around $\Gamma$=1.5. This is identical to what stellar mass X-ray binaries exhibit, indicating that there is probably an intrinsic switch in the accretion mode, from advection-dominated flows to standard (disk/corona) accretion modes.

\section{Sample Definition and Data Analysis}

Our sample has been obtained by cross-matching the SDSS DR7
spectroscopic sample with the Chandra Source Catalog.  We began with a
Bayesian-selected 2cross-match of the Chandra Source Catalog
(CSC Rev1.1; Evans et al. 2010) and the SDSS (York et
al. 2000), performed by the Chandra X-ray Center (CXC;
Rots et al. 2009), containing 16852 objects with both X-ray detections
and optical photometric objects in SDSS DR7.  Detailed visual
inspection of matches was performed to eliminate obviously saturated
optical sources, or uncertain counterparts in either band.
Since both redshifts and emission line measurements are
required for this study, the sample was further restricted to objects
for which there also exist SDSS optical spectra, leaving $\sim$2000
objects.

To take advantage of the diagnostic power of the H$\alpha$/[NII]
emission line complex, we set a limit of z$<$0.392 as for the
Constantin et al. (2009) sample, yielding 739 objects; of these, 685
are new relative to the aforementioned sample. The SDSS spectra for all 739
objects were downloaded and checked by eye to exclude objects with
serious artifacts in the spectrum or with grossly incorrect
redshifts. The latter included primarily stars and several broad
absorption line quasars. Upon completion, 714 spectra remained,
corresponding to 682 distinct objects.

A number of the objects are present in multiple Chandra obsids.
For simplicity, we merely selected the best observation to use in the
X-ray spectral fitting, primarily favoring the smallest off axis angle
(OAA, $\theta$) and the longest exposure time.  Upon further analysis
of the available X-ray data, we rejected 50 objects that were either
saturated, or too close to a chip edge. 

\subsection{Optical Spectroscopic Analysis} 

We limit the investigation to z$\lax$0.4 so that the $H{\alpha}$ and
other key emission lines are available within the wavelength range of
the SDSS spectrograph to perform classic emission line ratio
classification (e.g., Baldwin, Phillips \& Terlevich 1981 - BPT
hereafter; Kewley et al. 2006).  We fit optical spectra as described
in Zhou et al. (2006), 
beginning with starlight (using galaxy templates of Lu et al. 2006)
and nuclear emission (power-law) components that also account for
reddening, blended Fe\,II emission and Balmer continuum fitting.
Iterative emission line fitting follows, using multiple Gaussian or Lorentzian
profiles where warranted to fit broad and narrow line components.
Best template fits of the underlying host star light provide
estimates of the stellar mass, $M_{\rm BH}$ (via $\sigma_*$), along with mean
stellar ages via the strength of the 4000\AA-break and the H$\delta$
Balmer absorption line.

The optical spectral measurements include stellar velocity dispersions
with  errors for all galaxies as well as measurements of numerous
emission line fluxes. For the broad -line objects we measure the full
width at half maximum (FWHM) of the H$\beta$ emission
line. Additionally, the AGN flux at 5100\AA\, is calculated for
these objects, along with the AGN fraction of the total continuum. 
Unlike previous studies, this method enables us to use spectroscopic
analysis that is as uniform as possible for a diverse sample. 

For our broad line objects, black hole masses  have been
retrieved either  from Shen et al (2011), who have compiled
virial black hole mass estimates of all SDSS DR7 QSOs using
Vestergaard $\&$ Peterson (2006) calibrations for H$\beta$ and C$\,IV$
and their own calibrations for Mg$\,II$.  For other broad-line AGN
(BLAGN) - predominantly those spatially-resolved Sy\,1s that are not
targeted by the SDSS QSO programs - we use our own H$\beta$ emission
line fits. For all galaxies lacking broad emission
lines, we use our measurements of $\sigma_*$ to calculate M$_{\rm BH}$
values using the M-$\sigma$ relation of Graham et al. (2011).

The number of objects with successful spectral analysis
includes both broad and narrow emission line galaxies, totaling
617 in our final sample. Figure 1 shows the de-reddened 
$r$ band magnitude (SDSS modelMag) for the sample, plotted against
redshift. 

\subsection{Multi-wavelength Data} 

A prime advantage of our CSC/SDSS sample, in comparison to deeper
pencil-beam X-ray surveys, is its relatively shallow depth that allows
for easier source identification in other wavelengths. We have
cross-correlated our spectroscopic sample with publicly available
GALEX (DR6; Morrissey et al. 2007), UKIDSS (DR4; Lawrence et
al. 2007), 2MASS (Skrutskie et al. 2006), VLA (Becker et al. 1995) and WISE
(Wright et al. 2010) catalogs.  We have retrieved these catalogs using
the Virtual Observatory (VO) TOPCAT tool (Taylor 2005). Using
Monte-Carlo simulations and the Fadda et al (2002) method, we have
concluded that a search radius of 2.5 arcsec provides us with a $P(d)$
$<$ 0.02, where $P(d)$ is the Poisson probability of a GALEX source to
have a random association within a distance $d$, yielding an expected
rate of random associations of less than 5$\%$. The GALEX catalog
contains only sources that were detected at S/N $>$ 5 in at least one
of the NUV, FUV filters. All matches were then visually inspected to
remove any apparent spurious associations. We have adopted a similar
method for the catalogs at other wavebands.

\section{OPTICAL SPECTRAL CLASSIFICATION}

To classify emission line sources we use  BPT  diagnostic diagrams,
which employ four line flux ratios: [O\,III]5007/H$\beta$,
[N\,II]6583/H$\alpha$, [S\,II]6716,6731/H$\alpha$ and [O\,I]6300/H$\alpha$.  We only consider 
emission lines detected with at least 2$\sigma$ confidence. Following Kewley et
al. (2006) classification criteria, the emission line objects are
separated into Seyferts, LINERs, composite objects and star-forming
galaxies. A quite large (25$\%$) fraction of the emission-line objects
remains unclassified as their line ratios, although accurately
measured, do not correspond to a clear spectral type in the two
diagrams.  For the majority, while the [N\,II]/H$\alpha$
ratio shows relatively high, Seyfert like values, the corresponding 
[S\,II]/H$\alpha$ and [O\,I]6300/H$\alpha$ place them in the composite or star-forming objects
regime.  Thus, because the [S\,II] and [O\,I]  emission lines are better AGN
diagnostics than [N\,II], these systems are likely to be excluded from
the AGN samples selected via these classifications. As a consequence,
our samples based on the 6-line classification are small.

\begin{figure} 
\begin{center}
\begin{minipage}[c]{8.9cm}
       \includegraphics[width=1.0\textwidth]{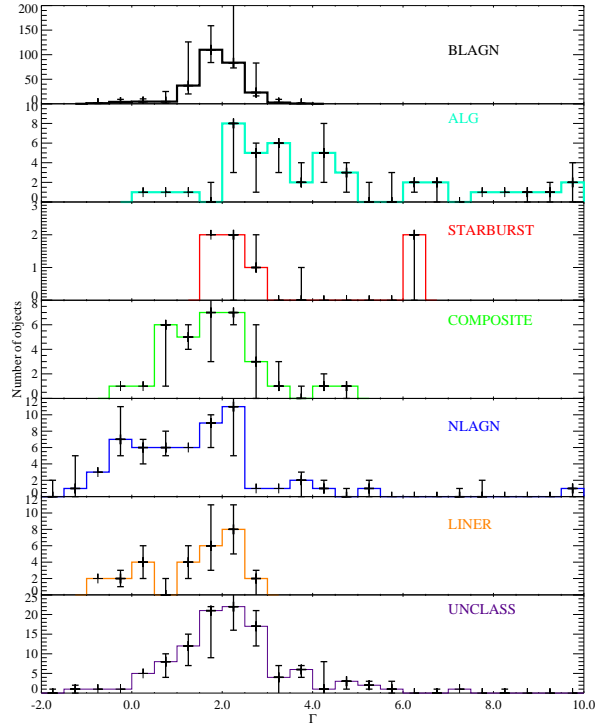}
\end{minipage}
\end{center}
    \caption
{The distribution of $\Gamma$ for the \logLx\,$>$42 sample,
  separated into subclasses based on 
  optical spectroscopic classification. Error bars show the Poisson
  errors on the number sources in each bin. From top to bottom:
  broad-line AGN, absorption line galaxies (ALG), star-forming
  objects, composite objects, narrow-line AGN and unclassified
  objects. }  
\label{GammaHistos}
\end{figure}

To enlarge our samples of galaxy nuclei of all spectral types, we also
explored an emission-line classification based on only the 
[O\,III]/H$\beta$ vs. [N\,II]/H$\alpha$ diagram, i.e., a 4-line
classification method, for the X-ray detected sources. The emission
line galaxy samples comprise thus all objects showing at least
2$\sigma$ confidence in the line flux measurements of these four lines
only. The delimitation criteria of star-forming and composite objects
remain unchanged, while Seyferts and LINERs are defined to be all
objects situated above the Kewley et al. (2006) separation line, and
with [O\,III]/H$\beta$ greater and less than  3 respectively.   Throughout the analysis presented in this paper we will call NLAGN the objects classified as Seyferts via the BPT diagrams.

\section{BOLOMETRIC LUMINOSITIES}

\begin{figure*} 
\begin{center}
\begin{minipage}[c]{8.9cm}     
        \includegraphics[width=1.0\textwidth]{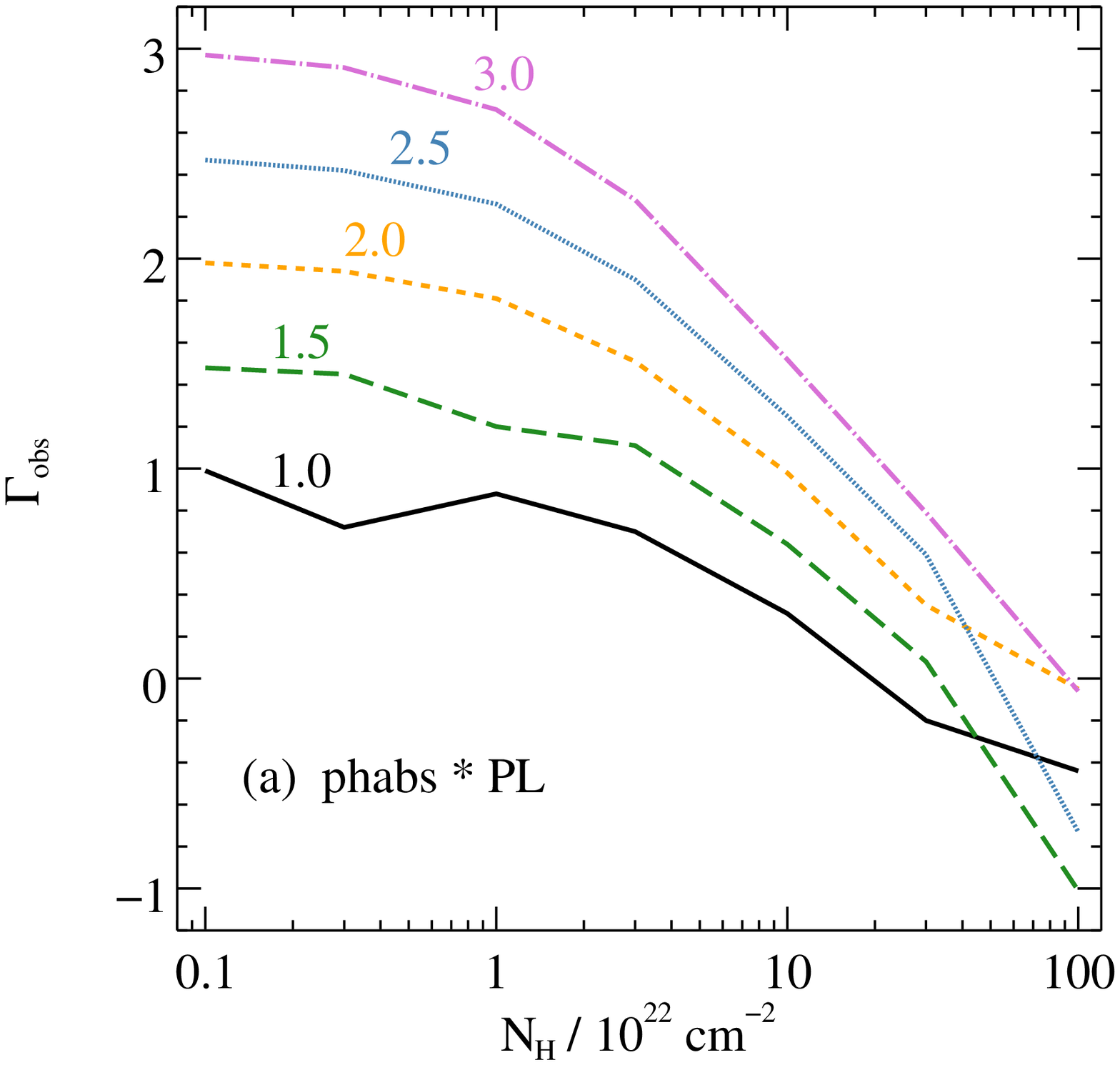} 
\end{minipage}
\begin{minipage}[c]{8.9 cm}
        \includegraphics[width=1.0\textwidth]{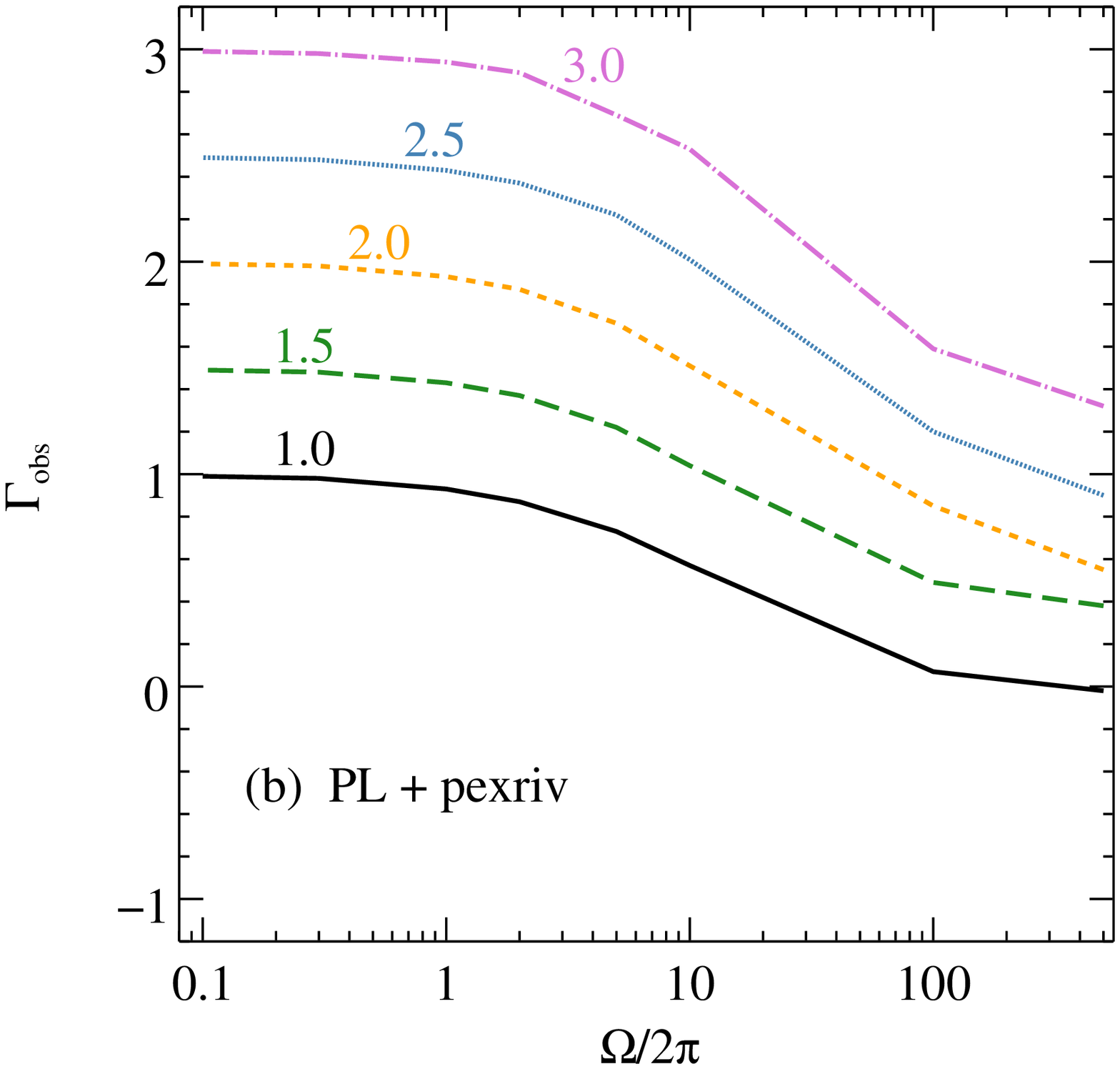}
        \end{minipage}
   \end{center}
\caption{Observed X-ray photon index as a function of the absorption
  column density ({\em LEFT:} XSPEC model phabs $\times$ power law) and
  reflection amplitude ({\em RIGHT:} XSPEC model pexriv with ionization
  parameter 10). The intrinsic X-ray photon index varies between 1.0
  and 3.0. X-ray power law spectra absorbed with N$_{H}$$>$10$^{23}$
  cm$^{-2}$, and reflection dominated X-ray spectra with
  $\Omega$/2$\pi$ $>$ 5-10 result in dramatically decreased observed
  X-ray photon index, reaching negative values for N$_{H}$$>$ few
  times 10$^{23}$ cm$^{-2}$. }
\label{obsGamma}
\end{figure*}
To estimate bolometric luminosities and check for the presence of
starburst and/or AGN activity in our sample, we fit the X-ray-to-radio
fluxes with various empirical SEDs of well-observed sources as
described in Trichas et al. (2012).  We have used a total of  41 such
templates, 16 from Ruiz et al (2010) and Trichas et al (2012) and 25 from Polletta et
al. (2006). We have adopted the model described in Ruiz et al. (2010) and Trichas et
al. (2012) which fits all SEDs using a $\chi$$^{2}$ minimization
technique within the fitting tool Sherpa (Freeman et al. 2001). Our
fitting allows for two additive components, using any possible combination
of AGN, starburst and galaxy templates. The SEDs are built  and
fitted in the rest-frame.  For each galaxy, we have chosen the fit
with the lowest reduced $\chi$$^{2}$ as our best fit model. Fractions
of AGN, starburst and galaxy contributions are derived from the SED
fitting normalizations  as these are derived from Ruiz et
al. (2010) model,  

\begin{equation}
 {F_{\nu}=F_{Bol}~(\alpha~u_{\nu}^{i}~+~(1-\alpha)~u_{\nu}^{j})}
\end{equation}

\noindent where $i$ and $j$ can be AGN, starburst or galaxy, F$_{Bol}$ is
the total bolometric flux, $\alpha$ is the relative contribution of
the i component to F$_{Bol}$ , F$_{\nu}$ is the total flux at
frequency $\nu$, while $u_{\nu}^{i}$ and $u_{\nu}^{i}$ are the
normalized i and j templates. 

\subsection{Comparison between SED fitting and optical spectroscopic
  classification} 

Among the 617 sources in our sample, 203 are broad-line AGN (BLAGN)
and 414 are narrow-emission/absorption line galaxies. The majority of the
sources are best fitted with a combination of templates however of the
203 BLAGN, in 168 (82$\%$) the dominant contribution is fitted with
one of our QSO templates, in 34 (17$\%$) with one of our NLAGN
templates and in only 2 ($<$1$\%$) with a non-AGN template. Of the 414
narrow emission or absorption line galaxies, in 399 (96$\%$) the dominant
contribution is fitted with one of our NELG/ALG
templates with only 15 (4$\%$) being fit with a BLAGN template.

Of the 414 narrow-emission/absorption line galaxies based on spectral
features and reliable emission line diagnostics, were possible, we
have 63 passives, 39 H$\,II$s, 77 transition objects, 130 Seyferts and
38 LINERs. In 84$\%$ of the passives the dominant contribution is
fitted by one of our elliptical templates, in 100\% of the H$\,II$s
with one of the star-forming templates and in 98\%  of the Seyferts
with one of our Seyfert templates. All (100\%) of the transition objects
require a combination of AGN and star-forming templates to fit
observed photometry. In the case of LINERs, the dominant contribution
is best fitted with a Seyfert, passive, or star-forming template in
37$\%$, 50$\%$ and 13$\%$ of the cases respectively.

Based on the above we can securely claim that the agreement of our SED
fitting with optical spectroscopic classification is excellent for all
types of these objects. 

\section{X-ray Spectral Fitting}

Based on the method used in Trichas et al. 2012, we perform X-ray
spectral fitting to all X-ray sources in our sample, using the CIAO
{\it Sherpa}\footnote{http://cxc.harvard.edu/sherpa} tool.  For each
source we fit three power-law models that all contain an appropriate
neutral Galactic absorption component frozen at the 21\,cm 
value:\footnote{Neutral Galactic column density $\nhgal$ taken from
Dickey et al. (1990) for the $Chandra$, aimpoint position on the sky.}
   (1) photon index $\Gamma$, with no intrinsic absorption component
(model ``{\tt PL}'')
   (2) an intrinsic absorber with neutral column $\nhintr$ at the
source redshift, with photon index frozen at $\Gamma=1.8$ (model
``{\tt PLfix}'').  Allowed fit ranges are $-1.5<\Gamma<3.5$ for {\tt
PL} and $10^{18}<\nhintr<10^{25}$ for {\tt PLfix}.
   (3) a two-parameter absorbed power-law where both $\Gamma$ and the
$\nhintr$ are free to vary within the above ranges while $\nhgal$ is fixed (model ``{\tt
PL\_abs}'').  All models are fit to the ungrouped data using Cash
statistics (Cash, 1979).  The latter model, {\tt PL\_abs},
is our default.

\begin{figure*} 
\begin{center}
\begin{minipage}[c]{8.9 cm}
        \includegraphics[width=1.0\textwidth]{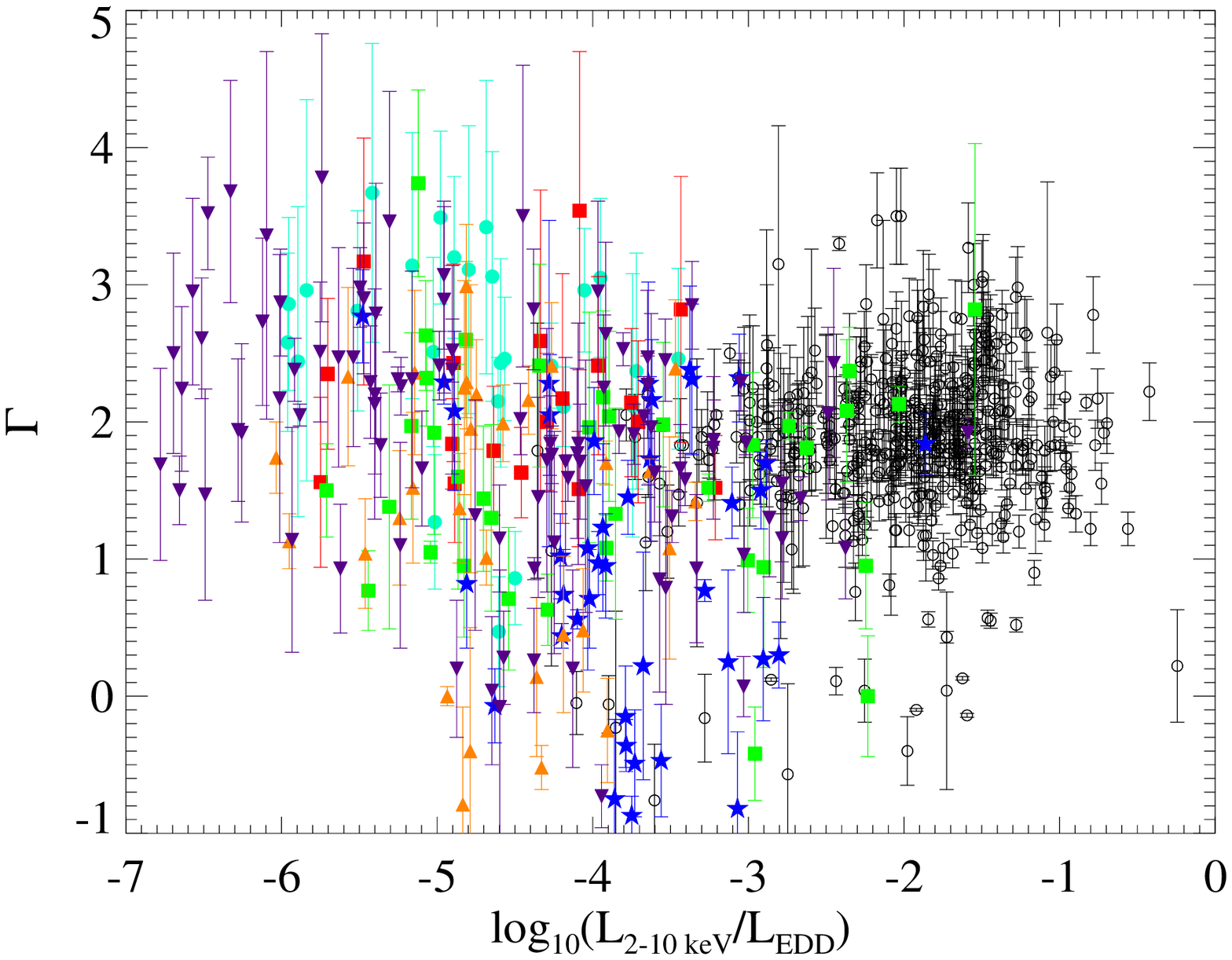}
        \end{minipage}
   \begin{minipage}[c]{8.9cm}     
        \includegraphics[width=1.0\textwidth]{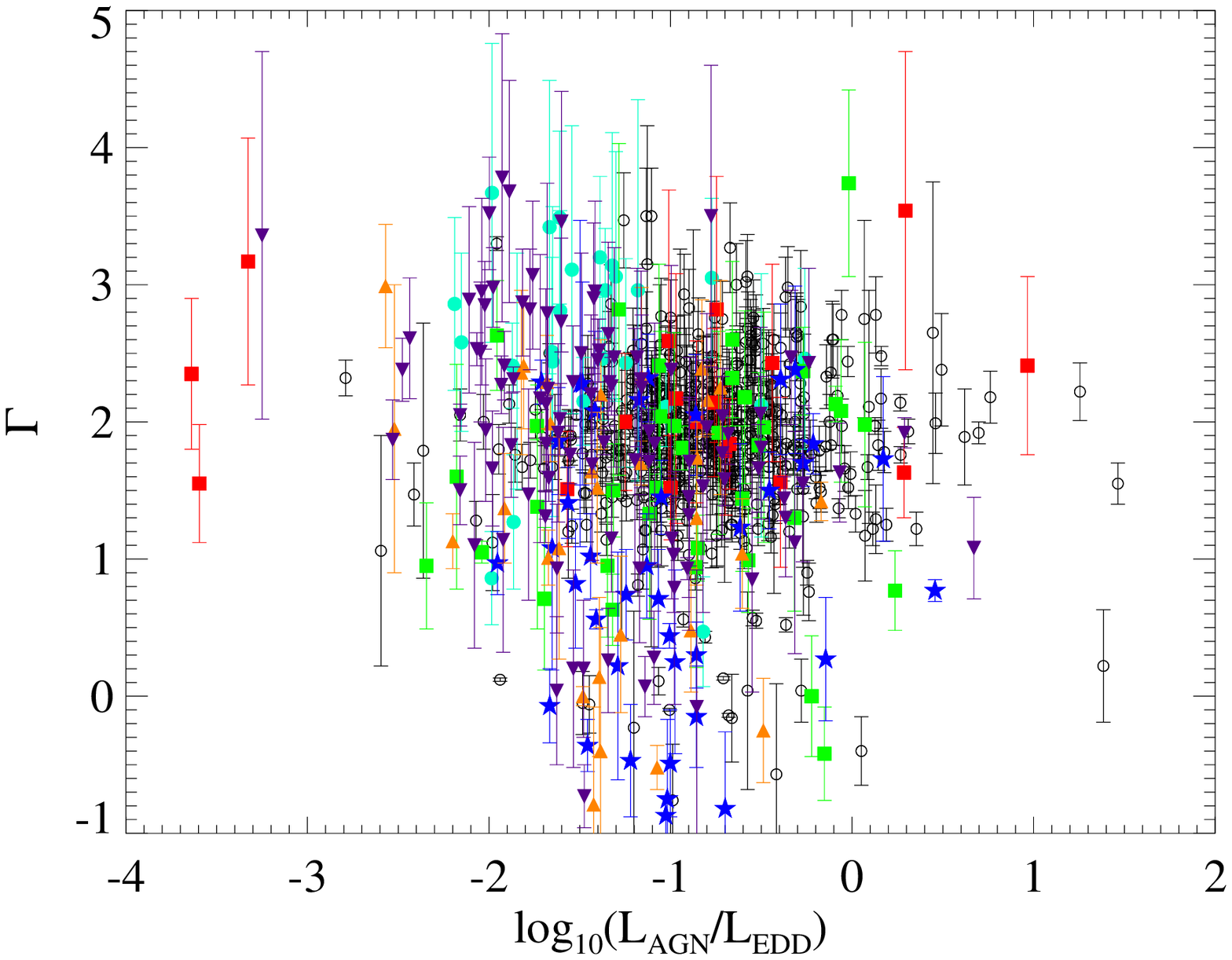} 
\end{minipage}
\end{center}
\caption{{\em LEFT:} $\Gamma$ - L$_{X(2-10 keV)}$/L$_{\rm Edd}$ relation
  for all z$<$0.4 CSC/SDSS AGN/Galaxies and ChaMP high-z QSOs (Trichas
  et al. 2012) with $\Gamma_{max}$ - $\Gamma_{min}$$\le$3 and minimum
  net counts of 20. Symbols are as described in Figure\,\ref{zr}. 
{\em RIGHT:} $\Gamma$ - L$_{AGN}$/L$_{\rm Edd}$
  relation for all z$<$0.4 CSC/SDSS AGN/Galaxies and ChaMP high-z QSOs
  (Trichas et al. 2012) with $\Gamma_{max}$ - $\Gamma_{min}$$\le$3 and
  minimum net counts of 20. L$_{AGN}$ is the bolometric AGN luminosity
  as calculated by Trichas et al. (2012) SED template fitting
  method.}
\label{GammaREdd}
\end{figure*}

As discussed in Trichas et al. (2012),
the best-fit $\Gamma$ from our default model is not correlated with
$\nhintr$, which illustrates that these parameters are fit with
relative independence even in low count sources.  Furthermore, the
best-fit $\Gamma$ in the default  {\tt PL\_abs} model correlates well
with that from the {\tt PL} model for the majority of sources; the
(median difference is 15\% of the median uncertainty).

A potentially useful Figure\,\ref{GammaHistos} shows the distribution
of $\Gamma$ for all sources with \logLx\,$>$42 divided by optical 
spectroscopic class.  As the X-ray emission in all these sources is
predominantly coming from an AGN, the peak of its
distribution appears to be at around $\Gamma$=2 as expected. This 
indicates that for luminous X-ray sources $\Gamma$ is not likely to be
severely affected by stellar X-ray emission from the host. However,
although the peak of each distribution is the same, the histogram
shape appears to change as we move to the type 2 spectral type sources corresponding to lower luminosity, or weaker accretion, e.g., LINERs and Composites, that could account for  different inclinations, and thus dustier circumnuclear regions and not necessarily for intrinsically hard ionizing continua, for which
there is a strong hard tail in the $\Gamma$ distribution and a sharp
drop above  $\Gamma$=2.  For ALG, the mode 
is at $\Gamma\sim 2.0$, confirming that these sources contain a
powerful AGN, but a a soft tail also indicates the likelihood of
either a different accretion mode, or perhaps contributions
from softer emission components such as thermal bremsstrahlung from
a hot interstellar medium, or even from circumgalactic hot gas, e.g.,
from the remnants of a 'fossil' galaxy group.  These sources are
typical examples of X-ray Bright Optically Inactive Galaxies (XBONG)
(Comastri et al. 2002). Four possible explanations have been proposed
for the nature of these objects (e.g. Green et al. 2004): a
$''$buried$''$ AGN (Comastri et al. 2002), a low luminosity AGN
(Severgnini et al. 2003), a BL Lac object (Yuan $\&$ Narayan 2004) and
galactic scale obscuration (Rigby et al. 2006; Civano et
al. 2007). The unclassified objects appear to follow a similar
distribution to the dustier objects. This is expected as these are
sources with very strong narrow emission lines which we fail to
classify because of the issues discussed in Section 3.

\begin{figure} 
\begin{center}
\begin{minipage}[c]{9.2cm}
       \includegraphics[width=1.0\textwidth]{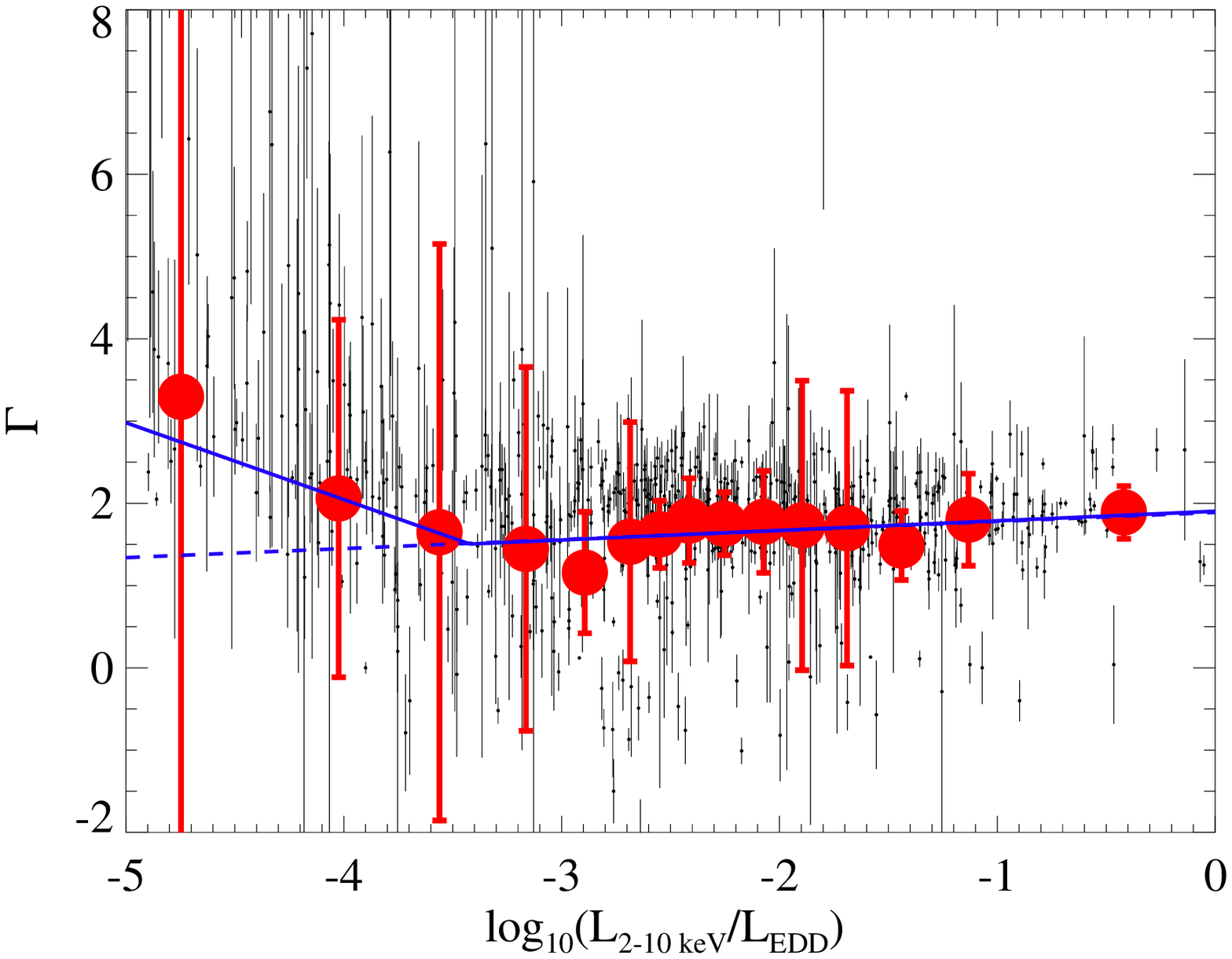}
\end{minipage}
\end{center}
    \caption{$\Gamma$, as a function of $L_{\rm 2-10keV}/L_{\rm edd}$
      ratio for all sources in our sample. Black dots represent the
      individual sources with their $\Gamma$ errors. The red circles
      represent the weighted mean values in each bin. The sample has
      been divided into 15 bins with the same number of sources per
      bin. The uncertainties shown represent the variance in the
      bin. The blue lines show the best fit with a broken linear
      model (solid line) and a single linear model (dashed
      line).}   

\label{avgGammaREddAll}
\end{figure}

\begin{figure*} 
\begin{center}
\begin{minipage}[c]{8.9 cm}
        \includegraphics[width=1.0\textwidth]{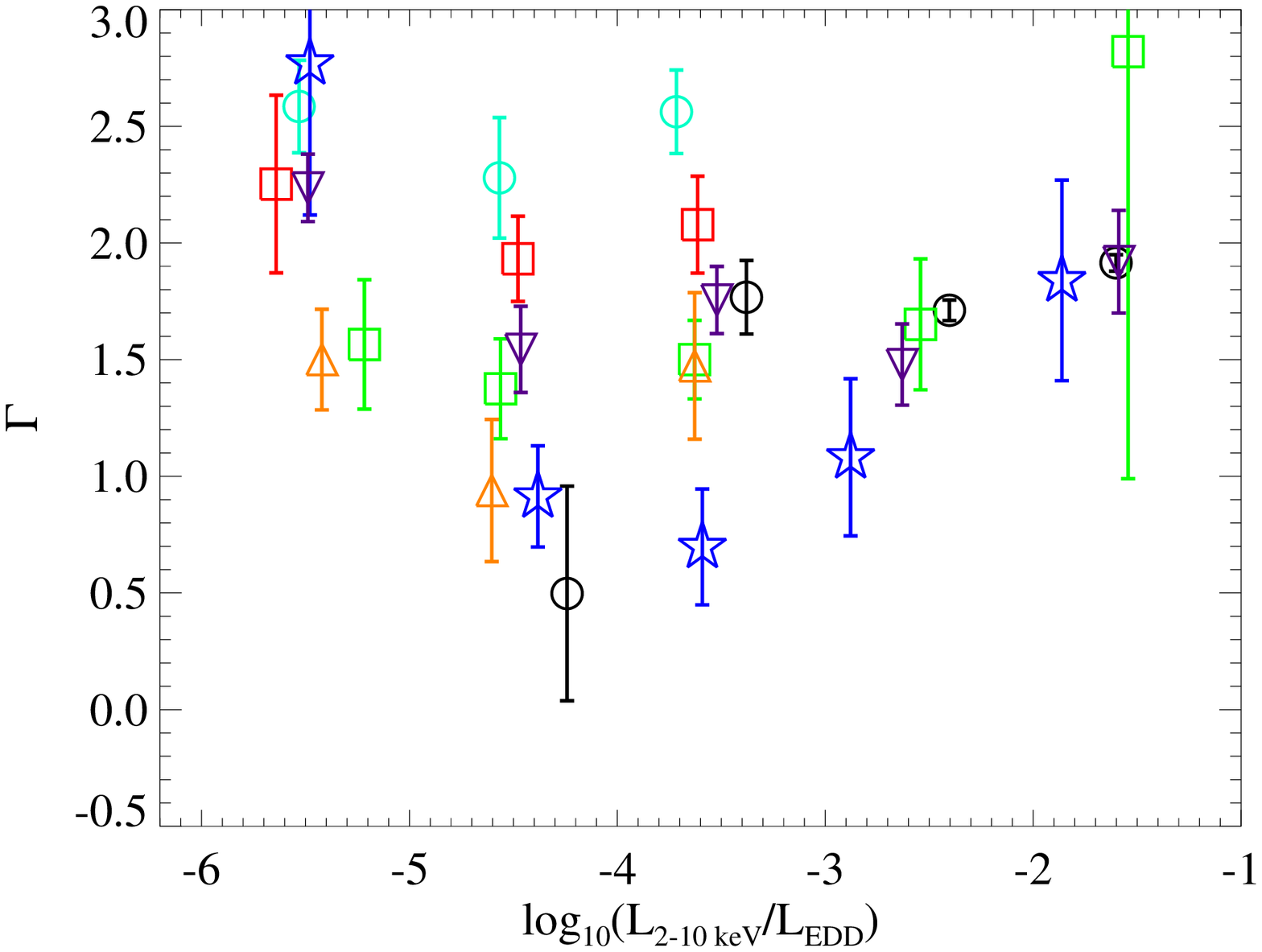}
        \end{minipage}
   \begin{minipage}[c]{8.9cm}     
        \includegraphics[width=1.0\textwidth]{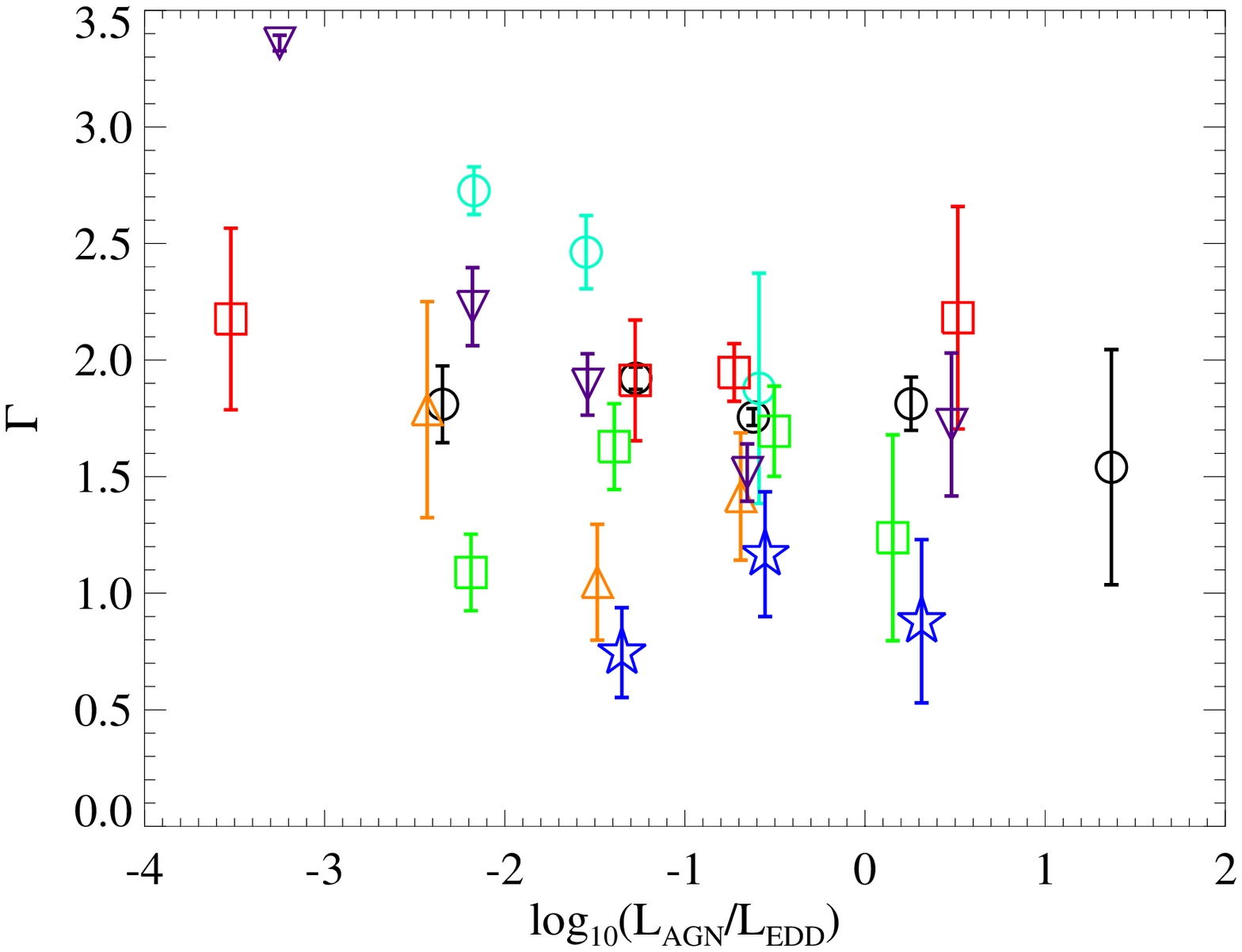} 
   \end{minipage}
\end{center}
\caption{Average $\Gamma$ per bin of $\Delta$log$\,\,L/L_{\rm Edd}$=1 for
 each spectral type of objects. Colors and shapes are the same as
 Figure\,\ref{GammaREdd}. {\em LEFT:} We have used the Eddington luminosity
 calculation with hard X-ray luminosity.  {\em RIGHT:} Eddington
 ratio using the SED-derived bolometric AGN luminosity.}
\label{avgGammaREdd}
\end{figure*}

For objects with sufficient X-ray counts and for which
none of the aforementioned models provide a satisfactory fit, multiple
additional models could be fitted to account for other possible sources
of X-ray emission, e.g., from the hot ISM, or a separate power-law
component from X-ray binary populations. In fact, most of our sources
have too few counts to warrant such detailed fitting. 

Objects for which we find very low (or even negative) $\Gamma$ could
be heavily intrinsically absorbed, in which case we observe primarily
the reflected component.  Modeling this in the 2-10\,keV band with a
power law would result in very hard apparently unphysical slopes.  We
show a simple simulation in Figure\,\ref{obsGamma} to illustrate. We
used two very simple XSPEC (Arnaud et al. 1996) models (1) phabs*PL
and (2) PL+pexriv (with ionization parameter 10, so effectively
reflection from neutral matter). The different contours show how the
measured $\Gamma$ depends on the (1) absorbing column $N_H$ and (2)
strength of reflection $\Omega/2\pi$, for intrinsic $\Gamma$ = 1.0,
1.5, 2.0, 2.5, 3.0.  These plots illustrate that negative observed
values of $\Gamma$ more likely correspond to the absorption case, at
least in this very simple approach.

To clarify this issue, deep X-ray exposures, preferably with hard
X-ray response extending above $\sim$8\,keV are required to allow more
detailed X-ray spectral analysis.  Additionally, we might expect that
such objects show less X-ray variability, since intrinsic variability
would be averaged by the reflection process (Sobolewska $\&$ Done 2007).

All multi-wavelength data are given in an online table available from the journal. A subsample of this table is given as an example in Table 1.

\section{$\Gamma$ - L/L$_{\rm Edd}$ relation for z$<$0.4 AGN/Galaxies}

The relation between the X-ray photon index $\Gamma$ and the Eddington
ratio for the entire SDSS/CSC sample of sources with optical spectra
at z$<$0.4 is illustrated in Figure\,\ref{GammaREdd}. X-ray luminosity
has been calculated using the method described in Green et al. (2011)
and bolometric luminosities have been calculated as described in
Section 4. Figure\,\ref{GammaREdd} shows 484
sources with minimum net counts of 20 and where the difference between
the upper 
and lower 90\% confidence limits to $\Gamma$ ($\Gamma_{\rm max} -
\Gamma_{\rm min}$) $\le$3. These selection criteria are applied in
order to include only sources that have meaningful X-ray spectral fits
(Section 5).  Different colors in Figure\,\ref{GammaREdd} represent
the different spectral classes as shown in Figure\,\ref{zr}.  To allow
sampling of higher accretion rates, the 
BLAGN sample in Figure\,\ref{GammaREdd} contains both CSC/SDSS
z$<0.4$ QSOs and high redshift QSOs from the Chandra Multwiavelength
Project (ChaMP) spectroscopic sample of Trichas et al. (2012). 

We have estimated the bolometric luminosity of the AGN component for
every SED in our sample for the purpose of testing whether its
relationship with Eddington ratio might reveal a more tightly
correlated trend, since the full AGN power is better estimated thereby.
However, the two panels in Figure\,\ref{GammaREdd}, which differ only
in the usage of $L_X$ vs L$_{AGN}$, reveal that $L_X$/L$_{\rm 
 Edd}$ results in a stronger trend, more easily separating the
AGN-dominated objects at higher Eddington ratio. 

We first suspected that the use of $\Gamma$ in the calculation of
$L_X$ itself might cause part of the correlation.
We therefore examined plots instead using a fixed $\Gamma$=1.9
to calculate $L_X$, and note no phenomenological difference
(the change is $\lax$0.1 in the Eddington ratio for $>90\%$ of the
objects).  There may be a failure of SED fitting to assess correctly
bolometric luminosities in cases where only a limited number of
photometric bands is available, or the reasons may be physical;
the relative optical-infrared contribution to $L_{\rm Bol}$
may be larger especially for star-forming NELGs or ALGs.  
We certainly expect - and we believe these plots confirm - that while
$L_X$ may provide an incomplete measure of AGN power, it is
the purest such measure available for a sample of this type.

To test whether we truly see an inflection point similarly to what is
observed in XRB we have selected all our objects with a clear
indication of AGN activity as identified in X-rays, namely sources
with \logLx\,$>$42. We have then split them in bins with equal number
of sources per bin and calculated average $\Gamma$ and $L_X$/L$_{\rm
  Edd}$ values. An ordinary least squares bisector method for linear
regression was used for the fitting. Figure\,\ref{avgGammaREddAll}
shows a very similar feature to that noted by Wu $\&$ Gu
(2008). The inflection point found by the latter is at the 
same $L_X$/L$_{\rm Edd}$ value as in our Figure\,\ref{avgGammaREddAll}
for AGN. 

A comparison between the $\chi$$^{2}$ values of the broken linear fit versus a single linear fit indicates that the broken linear fit is always a better fit. The latter as well as the inflection point value are independent from the number of bins used for the fitting. To test the statistical significance of the  $\chi$$^{2}$  results between broken and single linear fits, we have performed the p-value test (e.g. Sturrock $\&$ Scargle, 2009) that compares the likelihood ratios of fits done with a null hypothesis (i.e. single linear fit) and those with an alternative statistic (i.e. broken linear fit) using data simulated with Poisson noise. We have run the test using 20,000 simulations. The null hypothesis can be rejected if the p-value $\le$ 0.05 which is the default significance level for this kind of test. For our sample p-value $<$ 0.004 every time we run the test suggesting that the broken liner fit with a single inflection point is at 5$\%$ significance level, always a better solution to the single linear fit.

In Figure\,\ref{avgGammaREdd} we use the exact same selection sample
as in Figure 2 to plot the average $\Gamma$ per bin of log$\,\,L/L_{\rm
  Edd}$=1 separately for each spectral type. Number of objects differ
per bin which might affect the error calculation. While in the case of
the L$_{AGN}$/L$_{\rm Edd}$ the points show no trend, in the case of
$L_X$/L$_{\rm Edd}$ it is obvious that narrow-line AGN show a simila
break to the one observed in X-ray binaries (e.g. Wu $\&$ Gu 2008;
Sobolewksa et al. 2011). For other spectral types, no clear results
can be drawn as the populations do not cover the entire range of
L/L$_{\rm Edd}$.

\begin{figure*} 
\begin{center}
\begin{minipage}[c]{8.9 cm}
        \includegraphics[width=1.0\textwidth]{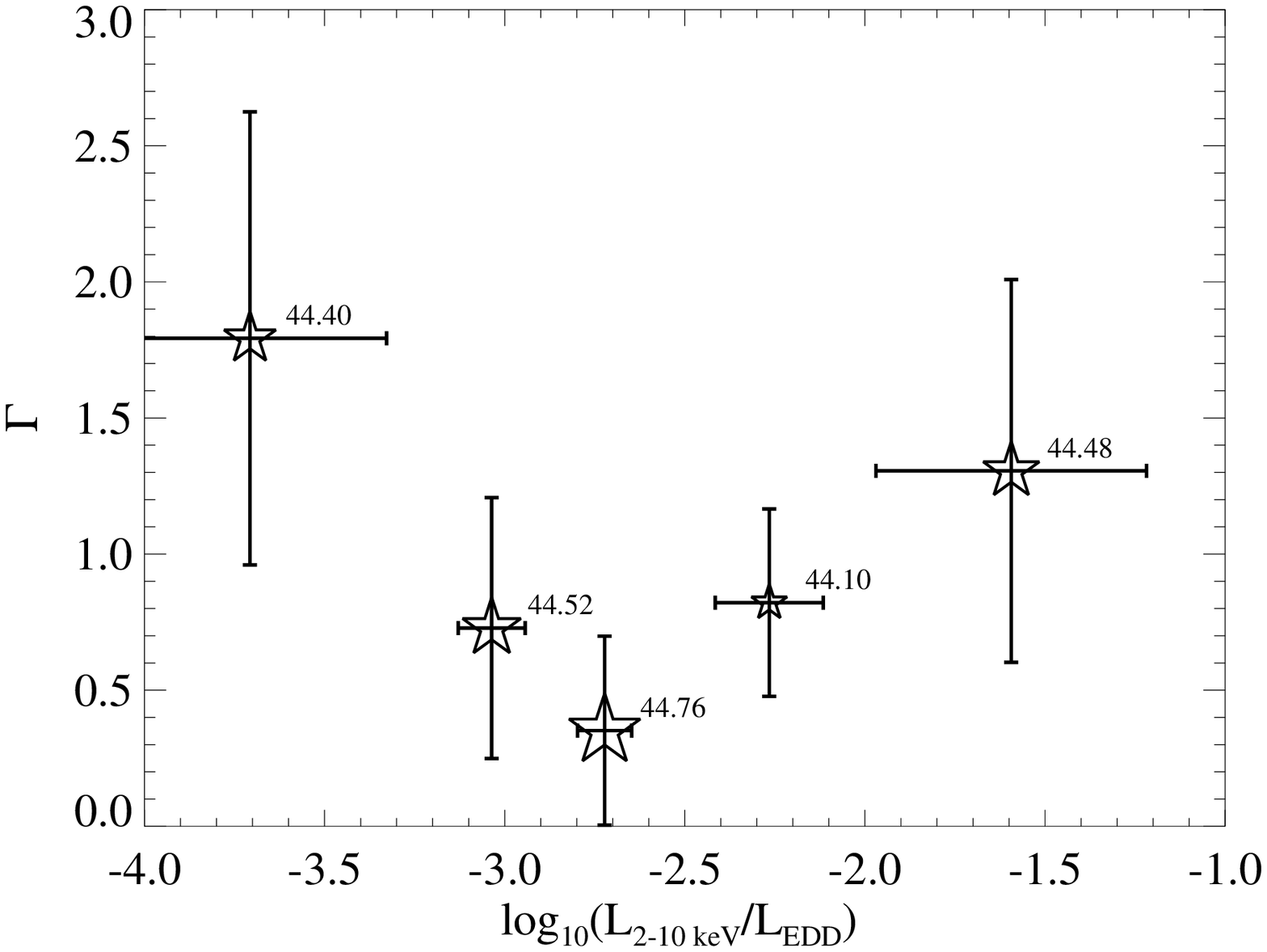}
        \end{minipage}
   \begin{minipage}[c]{8.9cm}     
        \includegraphics[width=1.0\textwidth]{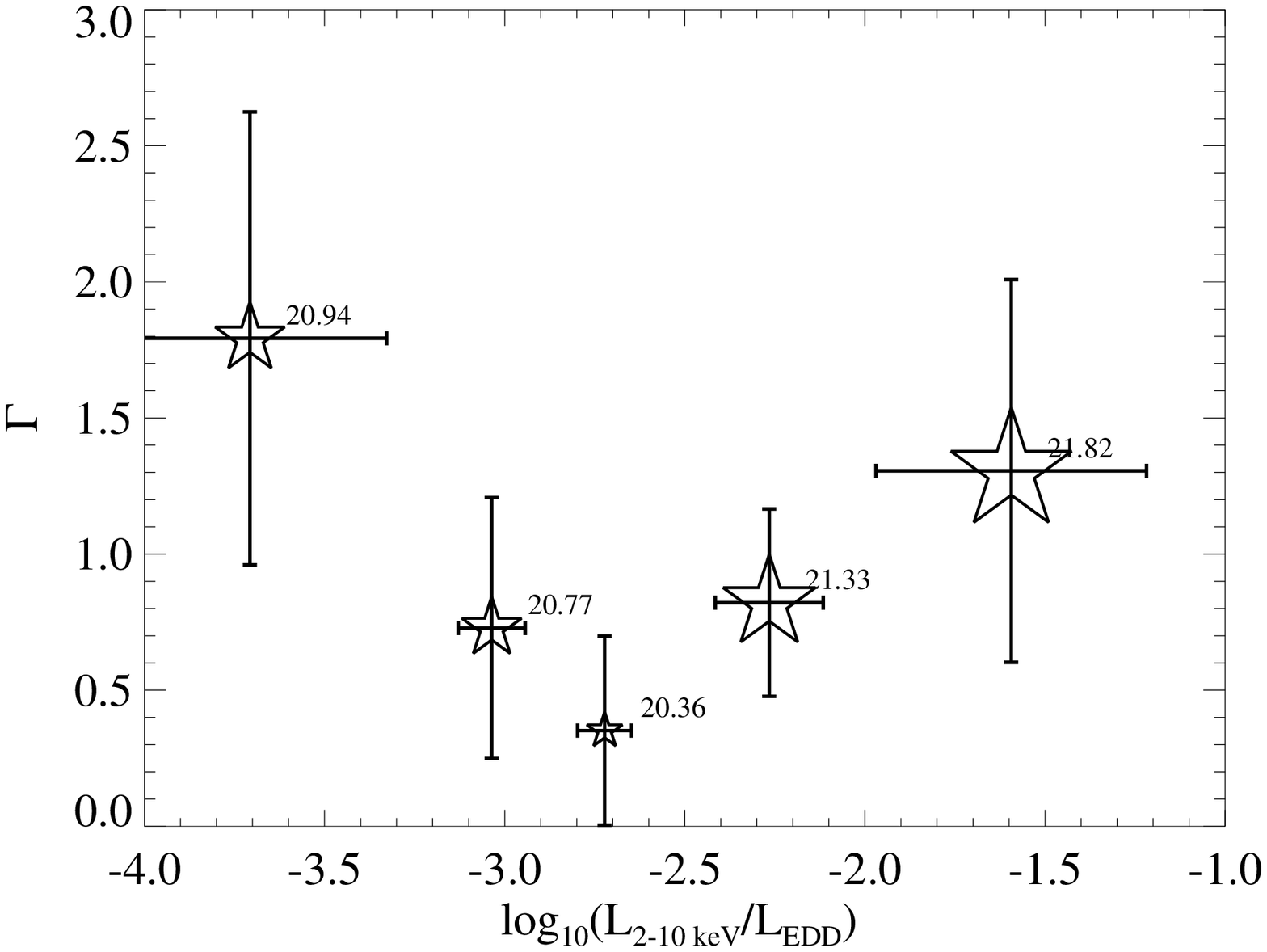} 
\end{minipage}
\end{center}
    \caption
{{\em LEFT:} Average $\Gamma$ versus Eddington ratio for all
narrow-line AGN with \logLx\,$>$42 erg/sec.  The stars
represent the mean $\Gamma$ values in each Eddington ratio bin. Each
bin contains the same number of sources. Error bars are the
1$\sigma$ $\Gamma$ error in each bin. The mean starburst luminosity is
given next of each star in erg/sec. {\em RIGHT:} Average
$\Gamma$ versus Eddington ratio for all narrow-line AGN with
\logLx\,$>$42. The stars represent the mean
$\Gamma$ values in each Eddington ratio bin. Each bin contains the
same number of sources. Error bars are the 1$\sigma$ $\Gamma$ error
in each bin. The mean number of N$_{H}$  is given next of each
star.} 
\label{avgGammaREddNLAGN}
\end{figure*}

 A closer inspection of each of the
spectral types for the entire CSC/SDSS spectroscopic z$<$0.4
population in Figure\,\ref{avgGammaREdd} indicates that the 
subsample that exhibits the strongest such trend is the population
of NLAGN (blue stars). We have tested whether there is an underlying
extrinsic cause for this strong trend for NLAGN mainly by checking
how the difference in column densities and star-formation properties
might affect it.  Figure\,\ref{avgGammaREddNLAGN} (left) shows these
trends for NLAGN with $log$ $L_{X(2-10 keV)}>42$.  Star-formation
values (the mean luminosity attributed to the best-fit star-forming
template components) have been retrieved by our SED fitting method. 
We might suspect that a significant contribution from star formation
might artificially soften our measured $\Gamma$.  However, 
the star-formation contribution from the host remains fairly constant
as a function of  $L_X$/L$_{\rm Edd}$ and in fact, the strongest
contribution is where measured mean $\Gamma$ is hardest, so
star-formation is not likely to affect the trend significantly.  
Another extrinsic effect might be that the hardest measured $\Gamma$
values around the inflection point could be caused by larger,
yet poorly-fit individual column densities, and/or 
some inverse correlation between the fit parameters $\Gamma$
and \nhintr\, as they compete to model the observed spectral shape. 
Figure\,\ref{avgGammaREddNLAGN} (right) shows some marginal
evidence that this could be a problem, because the hardest bin has the
lowest mean \nhintr. 

We retested our fits to the relations
$\Gamma$, as a function of $L_{\rm 2-10keV}/L_{\rm edd}$
for all sources in our sample after excluding all spectroscopically
identified narrow-line AGN, and find no significant difference.

\section{The \aox\ - L/L$_{\rm Edd}$ relation}
Trends of \aox\, vs. Eddington ratio for AGN are expected to 
be analogous to the SED variations of XRBs as well (e.g. Sobolewska et al.
2011). Indeed, such \aox\, trends may be expected to be stronger
than those with $\Gamma$, because the optical/UV emission tracks disk
emission with greater separation from AGN X-ray emission, which is
more strongly dominated by the corona.

\begin{figure} 
\begin{center}
\begin{minipage}[c]{8.9cm}
       \includegraphics[width=1.0\textwidth]{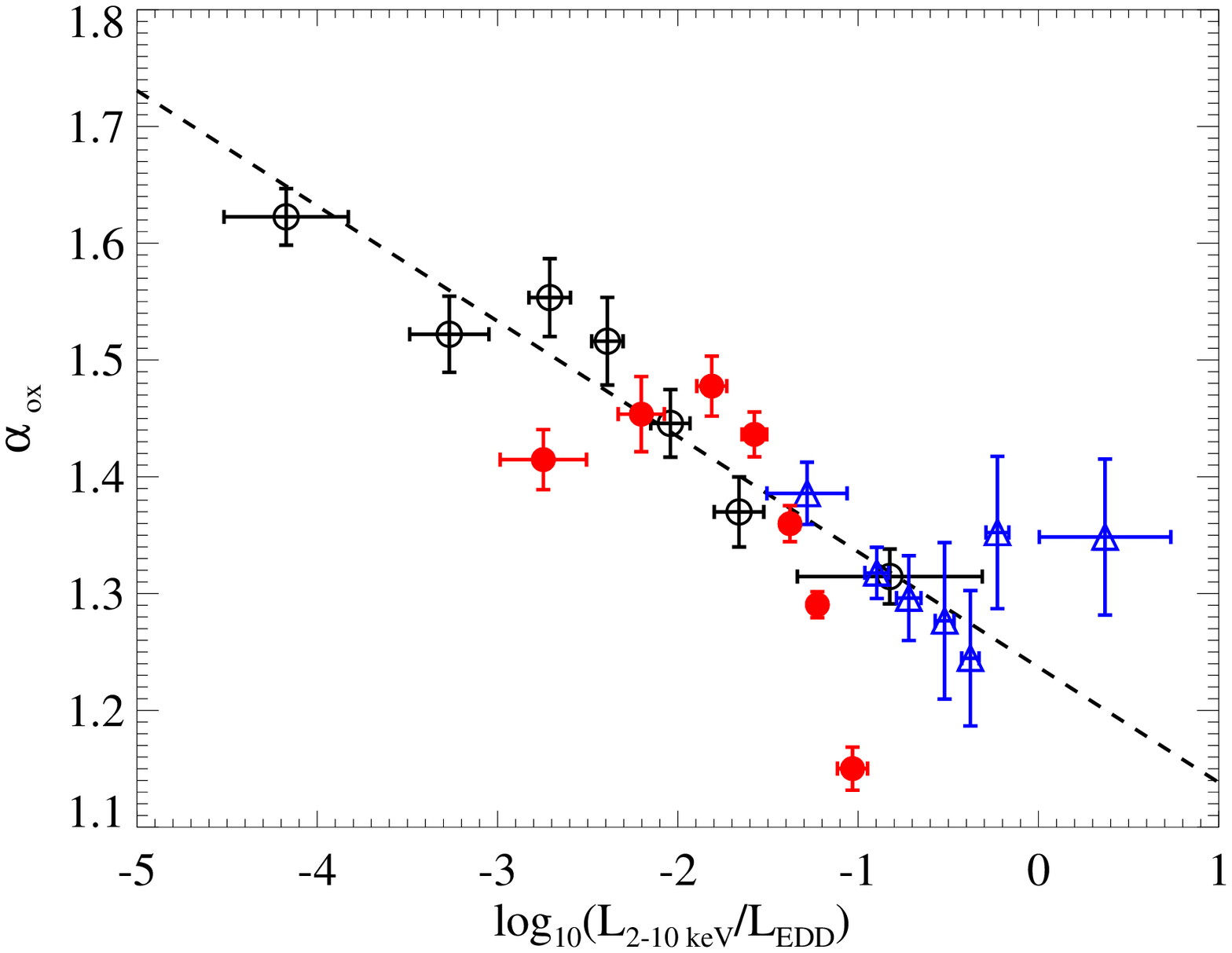}
\end{minipage}
\begin{minipage}[c]{8.9cm}
       \includegraphics[width=1.0\textwidth]{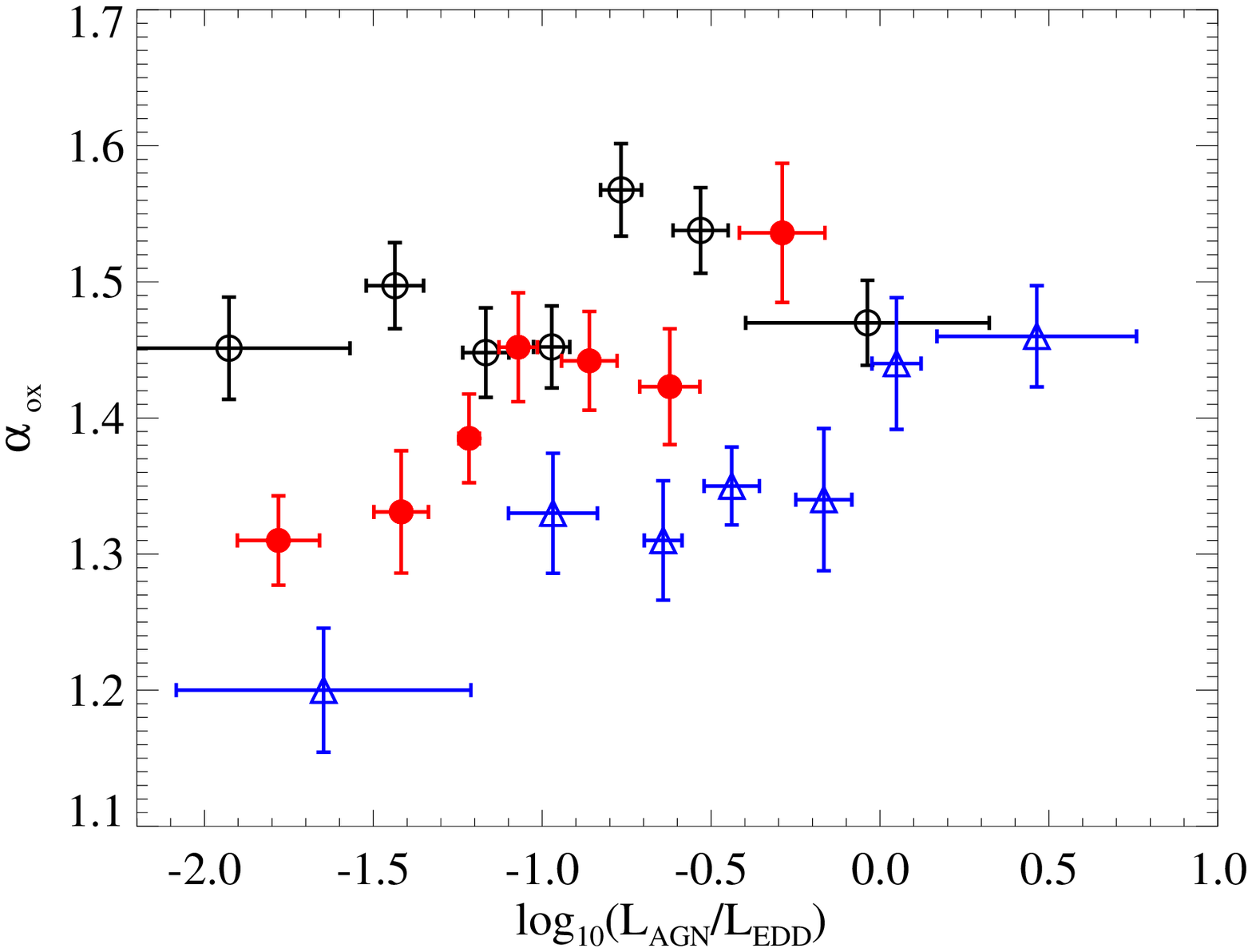}
\end{minipage}
\end{center}
    \caption{Average values of $\alpha_{\rm ox}$ vs 
$L_X/L_{\rm Edd}$ ratio for our entire sample with \logLx\,$>$42
erg/sec (open black circles).  Each bin contains the same number of
sources.  The best-fit OLS Y(X) regression plotted for our 
sample alone (dashed line).  We also show the sample of Grupe et
al. (2010; blue triangles), and of Lusso et al. (2010; filled red
circles), binned similarly. The apparently super-Eddington points from Grupe et al. (2010) are
probably influenced by uncertainties in the use of FWHM(H$\beta$)
to calculate M$_{BH}$ for these objects.}
\label{avgAOXREdd}
\end{figure}

Figure\,\ref{avgAOXREdd} shows the average values of \aox\, vs
$L_X$/L$_{\rm Edd}$ ratio for our sample of sources with \logLx\,$>$42
(open black circles).  Binned average points from Grupe et al (2010;
open blue triangles) and Lusso et al (2010; filled red circles) are
shown for comparison.   We estimate the 2-10\,keV luminosity from
the tablulated 0.--2\,keV luminosities in Grupe et al. (2010) by
simply assuming $\Gamma=1.9$.  The 150 
XMM-COSMOS BLAGN with M$_{\rm BH}$ estimates from Lusso et al (2010)
span a wide range of redshifts (0.2$<$z$<$4.25) and X-ray luminosities
between 42.4$<$log\,L$_{(2-10\,keV)}<$45.1.  However our study
extends the relationship down as far as log\,$L_X$/L$_{\rm Edd}$$<$-4.
At the high Eddington ratio end, there is a suggestion of an upturn,
as expected from the scaling experiments of Sobolewska et al. (2011).
However, the upturn is based on a small number of points from 
Grupe et al. (2010), and are suspect in any case, because
of the large uncertainties visible, and because log\,$L_X$/L$_{\rm
 Edd}$ exceeds unity. 

Since both referenced papers studied Eddington ratios using 
estimates of bolometric luminosity, we also compare using our $L_{\rm
 AGN}$ estimate from SED fitting. Here we see evidence
of a consistently positive trend.  The  Grupe et
al. (2010) points are again offset in the sense of being either
X-ray bright, or perhaps more likely suffering from underestimated
Eddington luminosities. 

\begin{figure} 
\begin{center}
\begin{minipage}[c]{8.9cm}
       \includegraphics[width=1.0\textwidth]{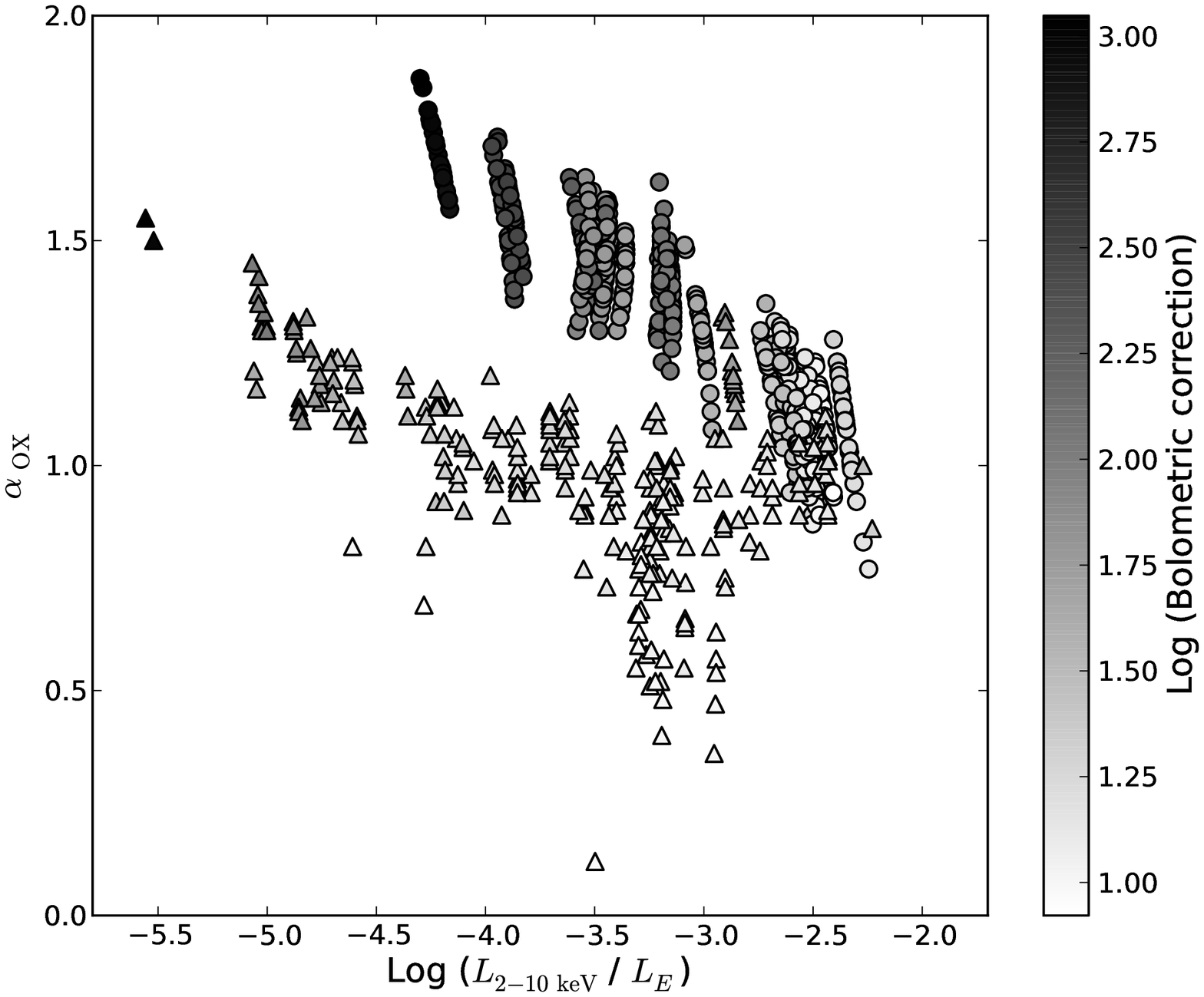}
\end{minipage}
\begin{minipage}[c]{8.9cm}
       \includegraphics[width=1.0\textwidth]{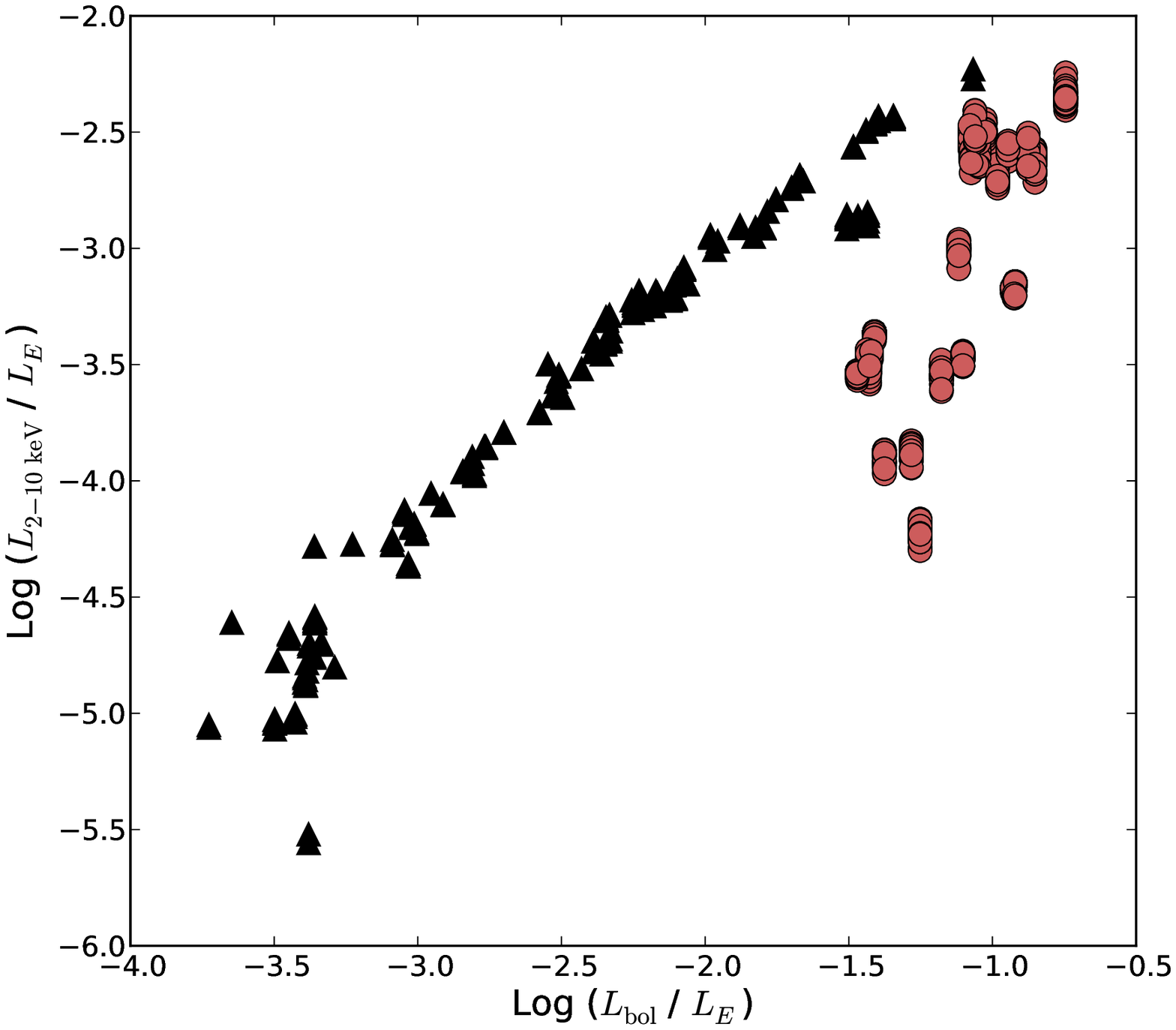}
\end{minipage}
\end{center}
    \caption{Simulations of the X-ray binary GRO\,J1655−40,
scaled to a simulated AGN population as described in the text. 
Circles show the  AGN-equivalent soft state, and triangles hard
state.  The bolometric correction from $L_X$ is shown as a grayscale
 (top).  In the bottom panel, Eddington ratios are plotted
against one another using  $L_X$ and $L_{\rm Bol}$.  The soft state
analogs (orange circles) show little correlation, because their SEDs
are so strongly dominated by disc (optical/UV) emission.}
\label{AOXsims}
\end{figure}

For comparison to Figure\,\ref{avgAOXREdd} (top), we offer 
Figure\,\ref{AOXsims}, based on the simulations of Sobolewska et
al. (2011).   Here, the observed SED evolution of 
the XRB GRO\,J1655−40 was scaled to a simulated population 
(BH mass distribution) of AGN, by stretching and scaling
a (3 vs. 20\,keV) disc-to-Componization index $\alpha^{\prime}$ 
to an AGN's analogous \aox\ (between 2500\AA\, and 2\,keV).
The exact values displayed are not of particular interest, because
the simulations convert expected analogous trends for AGN
extrapolating from the behavior in outburst of one particular X-ray
binary\footnote{For instance, the model XRB GRO\,J1655−40
was always below $\sim 0.2\,L_{\rm Bol}/L_{\rm Edd}$.} However, the
overall trends are instructive.   

In Figure\,\ref{avgAOXREdd} (top), 
the simulated AGN located on the upper branch (circles) correspond to
the soft state XRBs, with ultrasoft states located in the upper left
corner.  All points are shaded with the bolometric correction that
should be applied to the 2-10 keV luminosity to get $L_{\rm Bol}$.
The ultrasoft states have the largest bolometric correction 
because they are dominated by the accretion disk component with only
marginal contribution from the power-law tail.  These points most
probably correspond to X-ray weak AGN (.e.,g as indicated with a red
diamond in Fig\,3 of Vasudevan \& Fabian (2007).  

The simulated AGN sitting on the lower branch (triangles) correspond
to hard state XRBs. In general \aox\, is lower in the hard state than
in the soft state for comparable $L_X$/L$_{\rm Edd}$ due to less
vigorous accretion disk emission.   It is for these hard states that
we see not only a convincing anti-correlation of \aox\, with
 $L_X$/L$_{\rm Edd}$, but even an upturn just shy of the highest
Eddington ratios.  For the soft state (disc-dominated) AGN, changes in 
$L_X$ do not significantly change the $L_{\rm Bol}$, which results in
the varying bolometric correction seen at the top, and the poor
correlation between $L_{\rm X}/L_{\rm Edd}$ and $L_{\rm Bol}/L_{\rm Edd}$ 
seen in Figure\,\ref{avgAOXREdd}(bottom).  

We believe that the  AGN plotted in Figure\,\ref{avgAOXREdd}
may correspond to the soft state branch and bright (in terms of
$L_{\rm Bol}/L_{\rm Edd}$) part of the hard state branch, and that is 
why the turnover is weak in our observed sample.  This is confirmed in
the lower panel of Figure\,\ref{avgAOXREdd}, where $L_{\rm AGN}/L_{\rm Edd}$
is never lower than $\sim -2$.

\section{X-ray/Emission Line Relationships}

Observed relationships between emission line and X-ray luminosities
can be quite useful for predicting from optical ground-based
spectroscopy the (more expensive) X-ray exposure times required to
achieve a desired S/N. Such trends can also help us understand
the physical relationship between broad vs. narrow line emission 
and accretion power.

Panessa et al. (2006) studied various
correlations between X-ray and H$\alpha$, [O\,III] line luminosities,
from 60 ``mixed'' Seyferts, both narrow and broad lined Seyferts, in the Palomar survey of nearby 
($B_T<12.5$) galaxies (Ho, Filippenko, \& Sargent 1997)
and a sample of PG quasars from Alonso-Herrero et al. (1997).
Our sample extends to larger redshifts and X-ray luminosities
that may be better-matched to typical X-ray AGN studies.

\begin{figure*}
\begin{center}
\begin{minipage}[c]{8.9 cm}
        \includegraphics[width=1.0\textwidth]{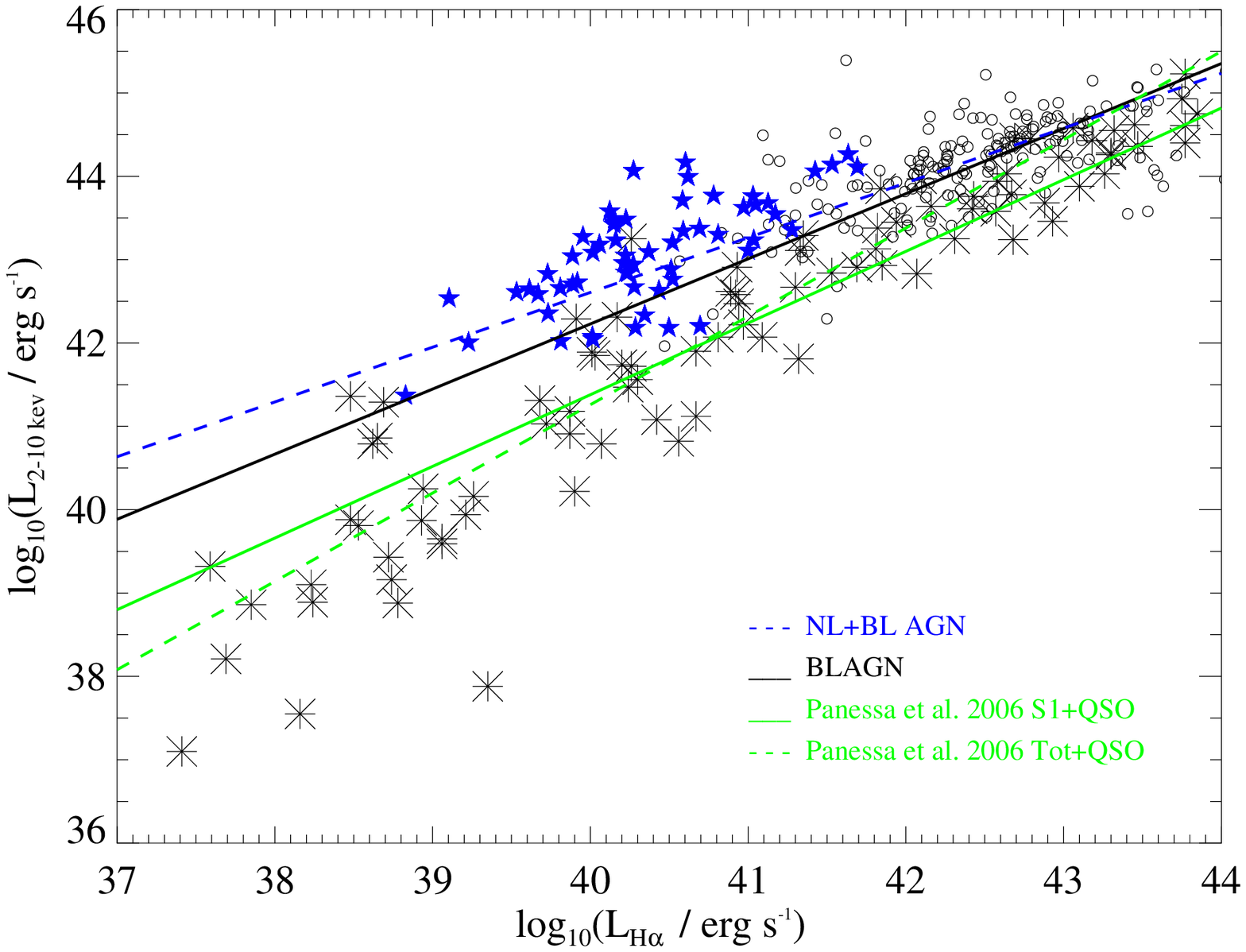}
        \end{minipage}
   \begin{minipage}[c]{8.9cm}     
        \includegraphics[width=1.0\textwidth]{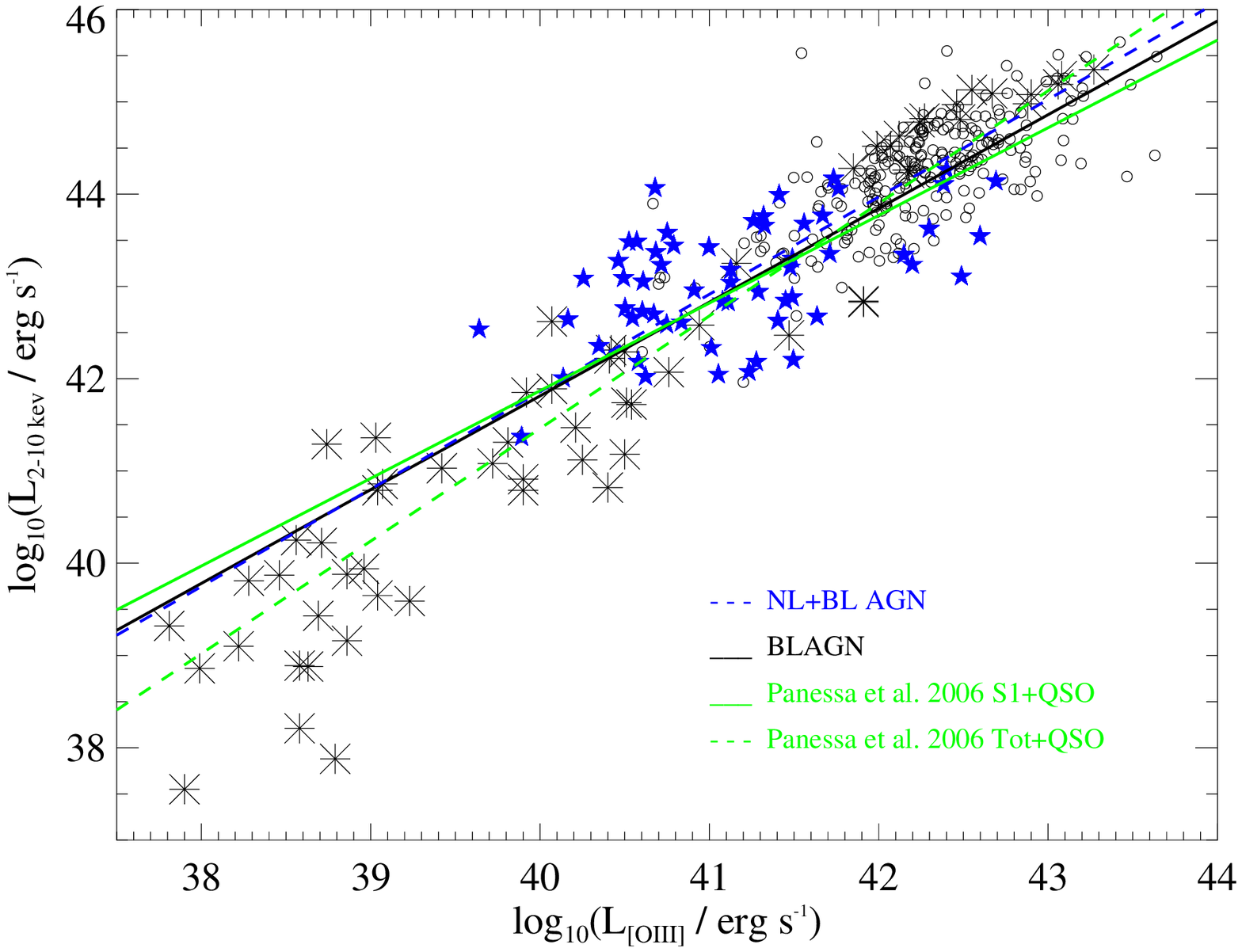} 
\end{minipage}
\end{center}
\caption{Logarithmic 2-10 keV luminosity versus logarithmic emission
  line luminosity.  {\em LEFT:}  For H$\alpha$ emission.
Open circles represent our BLAGN, blue
  filled stars are our NLAGN, and asterisks are Seyfert galaxies and
  low redshift QSOs from Ward et al. (1988) used in Panessa et
  al. (2006).  The black lines show the best fit (OLS) linear
  regression from our sample for BLAGN and NLAGN (solid) or just BLAGN
  (dashed).  The green solid line represents the best fit linear
  regression line from Panessa et al. (2006) for their full
  sample. 
{\em RIGHT:} For  [O\,III] emission. Same symbols as at left. 
 The black lines show the best fit (OLS) linear
  regression from our sample for BLAGN and NLAGN (solid) or just BLAGN
  (dashed).  The green lines represents the best fit linear
  regression line from Panessa et al. (2006) for Seyfert galaxies
(solid) and for their full sample (dashed), consisting of
bright type 1 Seyferts (Mulchaey et al. 1994) and a sample of PG
quasars (Alonso-Herrero et al. 1997).}  
\label{panessa}
\end{figure*}

Figure\,\ref{panessa} (left) shows the logarithmic 2-10 keV luminosity
versus  logarithmic H$\alpha$ luminosity. The black solid line shows
the best  linear (OLS) regression line 
   \begin{equation}
     {\log L_{X}=(1.02\pm 0.03)\log L_{H\alpha} - (1.18\pm 1.25)}
   \end{equation}
obtained by fitting our BLAGN sample (open black circles).
The green solid line (with slope $0.95\pm 0.07$ and intercept
$-3.87\pm 2.76$) represents the best-fit linear regression line
from Panessa et al. (2006) that they obtained by fitting the total
sample of Seyfert-1 galaxies and low redshift (PG) quasars.
While the difference in normalization is most apparent to the eye,
the values are consistent within the errors. 

Dashed blue and green lines represent the best fit to both broad and
narrow-line AGN for our sample and Panessa et al. 2006
respectively. Our fit is given by
     \begin{equation}
       {\log L_{X}=(0.66\pm0.03)\log L_{H\alpha} - (1.32\pm1.08)}
     \end{equation}
where Panessa et al. found slope $1.06\pm 0.04$ and intercept
$-1.14\pm 1.78$.  The offset between the samples is reminiscent of that
presented by Green, Anderson \& Ward (1992) in contrasting
the 60$\mu$m and X-ray emission between narrow- and broad- line
galaxies.  Both H$\alpha$ and (IRAS) 60$\mu$m luminosity is known to
correlate strongly with star-formation power, which may be relatively
strong in the nearby sample.

In the case of  the logarithmic 2-10 keV luminosity versus
logarithmic  [O\,III] luminosity for both BLAGN and NLAGN the agreement
is generally excellent.  We find 
    \begin{equation}
      {\log L_{X}= (1.26\pm0.04)\log L_{[OIII]} - (7.36\pm1.80)}
    \end{equation}
where the Panessa et al 2006 relationship  is fit with
slope $1.22\pm0.06$ and intercept $-7.34\pm2.53$.
We suggest that since [O\,III] is a higher ionization line, 
it tracks X-ray emission more accurately across a wider
range of accretion power, especially where the relative contribution
from star formation may be appreciable.

\section{Summary}

We confirm a significant V-shaped trend across a large
sample of X-ray detected galaxies and AGN in the plane of 
X-ray spectral hardness and Eddington ratio, when expressed
as $L_X$/L$_{\rm Edd}$.  The dispersion in the trend is significant
for a variety of reasons beyond intrinsic dispersion in
the accretion states of AGN.  The X-ray spectral fits are often
marginal due to poor photon statistics, and are generally unable
to model the known X-ray spectral complexity of real AGN,
including possible warm absorption, reflection components, etc.
We further acknowledge the rather significant uncertainties involved
in estimation of the Eddington luminosities from the $M_{\rm BH}$ -
$\sigma$ relation (for narrow emission and absorption line galaxies)
and the FWHM(H$\beta$)/$L_{\rm 5100\AA}$ method (for BLAGN).

Nevertheless, despite the intrinsic dispersion and measurement
uncertainties, we find on average a V-shaped correlation between X-ray
spectral hardness and Eddington ratio that is similar both in
shape and in the location of the inflection point, to analogous trends 
seen in X-ray binary systems as they vary. When separating AGN by
(optical spectroscopic emission line) classification, the strongest
trend is shown by the NLAGN.  A necessarily rather simple analysis
shows no evidence that this apparent trend is either strengthened 
or caused by the contributions of softer stellar emission components
or hardening due to intrinsic absorption.  We also test for a 
predicted V-shaped trend of X-ray to optical spectral slope \aox\,
with Eddington ratio, but find only a monotonic relationship
whereby BLAGN become relatively more X-ray bright (weak) 
compared to Eddington ratio expressed in terms of X-ray
(total AGN) luminosity.

\acknowledgments

Support for this work was provided by the National Aeronautics and
Space Administration through Chandra Awards Numbered
AR0-11018A and AR1-12016X, issued by the Chandra X-ray Observatory
Center, which is operated by the Smithsonian Astrophysical Observatory
for and on behalf of the National Aeronautics Space Administration
under contract NAS8-03060.  This research has made use of data
obtained from the Chandra Data Archive and the Chandra Source Catalog,
and software provided by the Chandra X-ray Center (CXC) in the
application packages CIAO, ChIPS, and Sherpa.  

Funding for the SDSS and SDSS-II has been provided by the Alfred
P. Sloan Foundation, the Participating Institutions, the National
Science Foundation, the U.S. Department of Energy, the National
Aeronautics and Space Administration, the Japanese Monbukagakusho,
the Max Planck Society, and the Higher Education Funding Council
for England. The SDSS Web Site is http://www.sdss.org/. 
The SDSS is managed by the Astrophysical Research Consortium for
the Participating Institutions. The Participating Institutions are
the American Museum of Natural History, Astrophysical Institute
Potsdam, University of Basel, University of Cambridge, Case
Western Reserve University, University of Chicago, Drexel
University, Fermilab, the Institute for Advanced Study, the Japan
Participation Group, Johns Hopkins University, the Joint Institute
for Nuclear Astrophysics, the Kavli Institute for Particle
Astrophysics and Cosmology, the Korean Scientist Group, the
Chinese Academy of Sciences (LAMOST), Los Alamos National
Laboratory, the Max-Planck-Institute for Astronomy (MPIA), the
Max-Planck-Institute for Astrophysics (MPA), New Mexico State
University, Ohio State University, University of Pittsburgh,
University of Portsmouth, Princeton University, the United States
Naval Observatory, and the University of Washington. 

\begin{sidewaystable*}
\caption{Sample of our full online catalog}
\begin{minipage}{35cm}
\begin{tabular}{lccccccccc}
\hline
  \multicolumn{1}{c}{SDSS~J name \footnote{SDSS identifier.}} &
  \multicolumn{1}{c}{redshift \footnote{Spectroscopic redshift.}} &
  \multicolumn{1}{c}{Class \footnote{Spectroscopic classification, 0: Passive Galaxy, 1: H II, 2: Transition Object, 3: Seyfert, 4: LINER, 5: unclassified, 6: broad-line AGN}} &
  \multicolumn{1}{c}{log\,$M_{\rm BH}$ \footnote{Logarithmic estimate of black hole mass (solar masses).}} &
  \multicolumn{1}{c}{net Counts \footnote{Number of counts for the 0.5-8 keV band.}} &
  \multicolumn{1}{c}{$\Gamma$ \footnote{Power-law slope from X-ray specrtal fitting with errors.}} &
  \multicolumn{1}{c}{${\rm N}^{intr}_{H}$\footnote{Best-fit intrinsic column density ${\rm N}^{intr}_{H}$.}} &
  \multicolumn{1}{c}{F(2-8 keV) \footnote{Hard (2-8 keV) X-ray flux in units of $10^{-14}{\rm erg ~s^{-1}~cm^{-2}}$ and errors.}} &
  \multicolumn{1}{c}{ log\,$L_X$/L$_{\rm Edd}$\footnote{Eddington ratio (2-8 keV (X-ray luminosity over bolometric luminosity).}} &
  \multicolumn{1}{c}{$L_{\rm AGN}$\footnote{AGN luminosity from best-fit SED in units of ${\rm erg ~s^{-1}}$.}} \\
\hline
\hline

J122137.2+295701 & 0.17 & 5 & 7.96$\pm 0.40$ & 26 & 1.84$^{+0.88}_{-0.58}$ & 21.04$\pm 0.52$ & 13.94$^{+6.37}_{-6.36}$ & -3.09 & 45.02\\
J123614.5+255022 & 0.18 & 5 & 7.86$\pm 0.42$ & 30 & 2.91$^{+1.66}_{-0.84}$ & 22.80$\pm 0.22$ & 11.53$^{+4.46}_{-4.43}$ & -3.01 & 45.09\\
J112314.9+431208 & 0.08 & 3 & 7.55$\pm 0.37$ & 123 & 1.86$^{+0.39}_{-0.36}$ & 21.94$\pm 0.37$ & 33.33$^{+3.21}_{-3.25}$ & -2.99 & 44.08\\
J153600.9+162839 & 0.38 & 6 & 7.83$\pm 0.22$ & 69 & 2.13$^{+0.24}_{-0.40}$ & 21.00$\pm 0.52$ & 139.4$^{+29.8}_{-30.0}$ & -1.13 & 45.51\\
J120100.1+133127 & 0.20 & 6 & 7.55$\pm 0.30$ & 68 & 1.68$^{+0.54}_{-0.47}$ & 21.77$\pm 0.03$& 203.5$^{+27.6}_{-28.0}$ & -1.32 & 44.99\\
J141652.9+104826 & 0.02 & 5 & 9.24$\pm 0.54$ & 200 & 2.05$^{+0.08}_{-0.08}$ & 20.60$\pm 0.30$ & 240.5$^{+5.6}_{-5.4}$ & -4.88 & 45.23\\
J090105.2+290146 & 0.19 & 3 & 8.14$\pm 0.34$ & 67 & 2.16$^{+0.44}_{-0.41}$ & 21.26$\pm 0.03$ & 43.95$^{+5.70}_{-5.63}$ & -2.62 & 45.11\\
J122959.4+133105 & 0.10 & 3 & 7.11$\pm 0.30$ & 228 & 0.3$^{+0.24}_{-0.22}$ & 21.28$\pm 0.15$ & 119.7$^{+8.2}_{-8.2}$ & -1.80 & 44.39\\
J122843.5+132556 & 0.25 & 0 & 9.00$\pm 0.31$ & 64 & 2.37$^{+0.86}_{-0.67}$ & 21.25$\pm 0.15$ & 137.9$^{+60.3}_{-59.61}$ & -2.72 & 45.30\\
J141531.4+113157 & 0.26 & 6 & 7.81$\pm 0.29$ & 293 & 1.74$^{+0.05}_{-0.05}$ & 20.48$\pm 0.01$ & 229.9$^{+4.4}_{-4.3}$ & -1.29 & 45.23\\
J111809.9+074653 & 0.04 & 2 & 7.26$\pm 0.49$ & 29 & 1.97$^{+0.68}_{-0.54}$ & 20.90$\pm 0.35$ & 4.38$^{+0.73}_{-0.73}$ & -4.17 & 44.42\\
J121531.2-003710 & 0.35 & 0 & 9.49$\pm 0.22$ & 21 & 3.67$^{+1.09}_{-0.87}$ & 21.94$\pm 0.15$ & 4.07$^{+1.67}_{-1.64}$ & -4.42 & 45.64\\
J145241.4+335058 & 0.19 & 6 & 7.92$\pm 0.30$ & 43 & 1.41$^{+1.05}_{-0.86}$ & 22.73$\pm 0.14$ & 53.84$^{+8.66}_{-8.57}$ & -2.33 & 44.96\\
J102451.2+470738 & 0.14 & 5 & 7.75$\pm 0.38$ & 57 & 2.25$^{+0.53}_{-0.48}$ &  21.23$\pm 0.01$ &  17.73$^{+2.56}_{-2.54}$ & -2.93 & 44.74\\
J082332.6+212017 & 0.02 & 1 & 7.64$\pm 0.42$ & 53 & 2.35$^{+0.55}_{-0.50}$  & 21.49$\pm 0.13$ &  17.13$^{+2.58}_{-2.60}$ & -4.70 & 42.14\\
J141910.3+525151 &  0.08 & 2 & 7.27$\pm 0.46$ &  43 & 1.44$^{+0.53}_{-0.40}$ &  20.95$\pm 0.55$ & 3.28$^{+0.57}_{-0.56}$ & -3.70 & 44.81\\
J011544.8+001400 &  0.04 & 3 & 6.38$\pm 0.50$ &  350 & 1.7$^{+0.17}_{-0.17}$ &  20.90$\pm 0.01$ & 100.10$^{+5.7}_{-5.73}$ & -1.89 & 44.26\\
J011522.1+001518 &  0.39 & 5 & 9.24$\pm 0.25$ &  386 & 1.31$^{+0.19}_{-0.19}$ & 22.25$\pm 0.30$  & 147.70$^{+7.80}_{-7.70}$ & -2.49 & 45.69\\
J082001.8+212107 & 0.08 & 1 & 6.53$\pm 0.42$ &  29 & 2.82$^{+0.97}_{-0.48}$ & 20.30$\pm 0.08$ & 11.03$^{+2.23}_{-2.21}$ & -2.43 & 43.93\\
J020925.1+002356 &  0.06 & 2 & 6.76$\pm 0.44$ &  40 & 1.83$^{+0.53}_{-0.34}$ &  20.30$\pm 0.01$ & 99.61$^{+18.99}_{-18.80}$ & -1.97 & 44.39\\
J153311.3-004523 &  0.15 & 3 & 8.01$\pm 0.28$ &  37 & 1.92$^{+1.27}_{-1.13}$ &  22.93$\pm 0.28$ & 164.8$^{+28.0}_{-28.4}$ & -2.16 & 44.82\\
J002253.2+001659 & 0.21 & 5 & 7.63$\pm 0.34$ &  150 & 1.03$^{+0.42}_{-0.40}$ & 22.55$\pm 0.52$ & 43.03$^{+3.69}_{-3.66}$ & -2.03 & 44.79\\
J155627.6+241800 & 0.12 & 2 & 7.38$\pm 0.37$ &  94 & 0.94$^{+0.73}_{-0.67}$ & 22.61$\pm 0.30$ & 119.1$^{+15.6}_{-15.5}$ & -1.90 & 44.66\\
J103515.6+393909 & 0.11 & 3 & 6.24$\pm 0.34$ & 33 & 1.7$^{+1.23}_{-1.01}$ & 22.83$\pm 0.23$ & 109.2$^{+20.2}_{-19.08}$ & -0.89 & 44.27\\

\hline\end{tabular}
\end{minipage}
\end{sidewaystable*}

\end{document}